\documentclass[10pt]{iopart}
\usepackage{iopams}
\usepackage{epsfig,epsf,graphics}
\usepackage{amssymb}
\usepackage{txfonts}
\usepackage{rotating}  
\def\rmr{{\rm r}}
\def\rma{{\rm a}}
\def\rms{{\rm s}}
\def\rmp{{\rm p}}

\def\rme{{\rm e}}

\def\rmS{{\rm S}}
\def\rmP{{\rm P}}

\def\rmH{{\rm H}}

\def\orb[#1 #2]{{$#1^{#2}$}}
\def\ls[#1 #2]{{$^{#1}${#2}}}
\def\tm[#1 #2 #3]{{$^{#1}${#2}$_{#3}$}}
\def\fl[#1 #2]{{#1}$\pm${#2}}

\def\rma{{\rm a}} 
\def\rmr{{\rm r}} 
\def\rme{{\rm e}} 
\def\rms{{\rm s}} 
\def\rmp{{\rm p}}

\def\rmS{{\rm S}} 
\def\rmP{{\rm P}}

\DeclareMathAlphabet{\scriptnew}{U}{eus}{m}{n}
\begin{document}
\bibliographystyle{plain}
%\setlength{\parindent}{3em}

%\thesaurus{02         % A&A Section 12: Atomic, molecular and nuclear data
%               (02.01.4;  % Atomic processes
%               02.03.2)} % Atomic data

\title{ Ionisation state, excited populations and emission of impurities in dynamic finite density plasmas.
\newline 
I: The generalised collisional-radiative model for light elements.} 

   \author{H P Summers\dag , W J Dickson\dag , M G O'Mullane\dag , N R Badnell\dag , A D Whiteford\dag , 
           D H Brooks\ddag , J Lang\S , S D Loch\$ and D C Griffin\#}

   \address{\dag\ Department of Physics, University of Strathclyde, 107 Rottenrow, Glasgow G4 0NG, UK}
   
   \address{\ddag\ George Mason University, 4400 University Drive, Fairfax, Virginia, VA 22030 USA}
   
   \address{\S\ Rutherford Appleton Laboratory, Chilton, Didcot, Oxon.  OX11 0QX, UK}
   
   \address{\$\ Department of Physics, Auburn University, Alabama, AL 36849, USA}
   
   \address{\#\ Rollins College, Winterpark, Florida, FL 32789, USA}

%\offprints{H.P. Summers}
\date{Received \today}

\markboth{H.P. Summers et al. Ionisation state. I.}
{H.P. Summers et al. Ionisation state. I.}

\begin{abstract}
 The paper presents an integrated view of the population structure and its role in establishing the ionisation
state of light elements in dynamic, finite density, laboratory and astrophysical plasmas.  There are four main issues,
the generalised collisional-radiative picture for metastables in dynamic plasmas with Maxwellian free electrons and its
particularising to light elements, the methods of bundling and projection for manipulating the population equations, the
systematic production/use of state selective fundamental collision data in the metastable resolved  picture to all
levels for collisonal-radiative modelling and the delivery of appropriate derived coefficients for experiment analysis. 
The ions of carbon, oxygen and neon are used in illustration.  The practical implementation of the methods described
here is part of the ADAS Project.
\end{abstract}  


%\keywords{atomic database -- electron collision rates --
%fine-structure transitions}

\maketitle
%
%________________________________________________________________

\section{Introduction}
\label{sec:sec1}

The broad mechanism for radiation emission from a hot tenuous plasma is simple.  Thermal 
kinetic energy of free electrons in the plasma is transferred by collisions to the internal energy of 
impurity ions,
\begin{eqnarray}
\label{eqn:eqn1}
	\scriptnew{A}+e \rightarrow \scriptnew{A}^* + e
\end{eqnarray}
where $\scriptnew{A}^*$ denotes an excited state and $\scriptnew{A}$ the ground state of the impurity ion.  This energy
is  then radiated as spectrum line photons which escape from the plasma volume
\begin{eqnarray}
\label{eqn:eqn2}
	\scriptnew{A}^* \rightarrow \scriptnew{A} + h\bar{\nu} .
\end{eqnarray}
where $h\bar{\nu}$ is the emitted photon energy and $\bar{\nu}$ its frequency.  Similarly ions in general increase or decrease their charge
state by collisions with electrons
\begin{eqnarray}
\label{eqn:eqn3}
\scriptnew{A} + e   & \rightarrow & \scriptnew{A}^+ + e + e	\nonumber \\
\scriptnew{A}^+ + e & \rightarrow & \scriptnew{A} +h\bar{\nu}
\end{eqnarray}
where $\scriptnew{A}^+$ denotes the next ionisation stage of impurity ion  $\scriptnew{A}$.  The situation is often referred to as the {\it
coronal picture}. The coronal picture has been the basis for the description of impurities in fusion plasmas for many years. However, the
progress towards ignition of fusion plasmas and to higher density plasmas requires a description beyond the coronal approximation.  Models of
finite density plasmas which include some parts of the competition between radiative and collisional processes are loosely called {\it
collisional-radiative}.  However, collisional-radiative theory in its origins (Bates \etal, 1962) was designed for the description of dynamic
plasmas and this aspect is essential for the present situations of divertors, heavy species, transport barriers and transient events.  The
present work is centred on {\it generalised collisional radiative} ($\scriptnew{GCR}$) theory (McWhirter and Summers, 1984) which is developed in
the following sections.  It is shown that consideration of relaxation time-scales, metastable states, secondary collisions etc. - aspects
rigorously specified in collisional-radiative theory - allow an atomic description suitable for modelling the newer areas above.   The detailed
quantitative description is complicated because of the need to evaluate individually the many controlling collisional and radiative processes, a
task which is compounded by the variety of atoms and ions which participate.  The focus is restricted to plasmas which are optically thin and not
influenced by external radiation fields, and for which ground and metastable populations of ions dominate other excited ion populations.  The
paper provides an overview of key methods used to expedite this for light elements and draws illustrative results from the ions of carbon, oxygen
and neon. The paper is intended as the first of a series of papers on the application of collisional-radiative modelling in more advanced plasma
scenarios and to specific important species.    

The practical implementation of the methods described here is part of the  ADAS (Atomic Data and Analysis
Structure) Project (Summers, 1993, 2004).  Illustrations are drawn from ADAS codes  and the ADAS fundamental and
derived databases.      

\subsection{Time constants}
\label{sec:sec1.1}

The lifetimes of the various states of atoms, ions and electrons in a plasma to radiative or collisional processes vary
enormously.  Of particular concern for spectroscopic studies of dynamic finite density plasmas are those of translational states of free
electrons, atoms and ions and internal excited states (including states of ionisation) of atoms and ions.  These lifetimes
determine the relaxation  times of the various populations, the rank order of which, together with their values relative to
observation times and plasma development times determines the modelling approach.  The  key lifetimes divide into two
groups.  The first is the {\it intrinsic} group, comprising purely atomic parameters, and includes metastable radiative
decay, $\tau_m$, ordinary excited state radiative decay $\tau_o$  and auto-ionising  state decay (radiative and Auger), 
$\tau_a$. The intrinsic group for a particular ion is generally ordered as
\begin{eqnarray}
\label{eqn:eqn4}
	\tau_a << \tau_o << \tau_m
\end{eqnarray}
with typical values
\begin{equation}
\label{eqn:eqn5}
  \tau_m \sim 10^1/z^8~s,\tau_o \sim 10^{-8}/z^4~s,\tau_a \sim 10^{-13}~s. 	
\end{equation}
where $z$ is the ion charge.  The second is the {\it extrinsic} group, which depends on plasma conditions - especially particle
density.  It includes free particle thermalisation (including electron-electron  $\tau_{e-e}$, ion-ion  $\tau_{i-i}$ and ion-electron
$\tau_{i-e}$),  charge-state change (ionisation $\tau_{ion}$ and recombination $\tau_{rec}$) and redistribution of population amongst
excited ion states ($\tau_{red}$).    
The extrinsic group is ordered in general as
\begin{eqnarray}
\label{eqn:eqn6}
	\tau_{ion,rec}  >> \tau_{i-e} >> \tau_{i-i}>>\tau_{e-e}
\end{eqnarray}
with approximate expressions for the time constants given by 
\begin{eqnarray}
\label{eqn:eqn7}
\tau_{rec} & \sim & [10^{11}-10^{13}](1/(z+1)^2)(kT_e/I_H)^{1/2}(\rm{cm}^{-3}/N_e)~~s	\nonumber\\ 
\tau_{ion} & \sim & [10^5-10^7](z+1)^4(I_H/kT_e)^{1/2}e^{\chi/kT_e}(\rm{cm}^{-3}/N_e)~~s	\nonumber\\ 
\tau_{i-i} & \sim & [7.0\times10^7](m_i/m_p)^{1/2}	\nonumber\\
& & (kT_e/I_H)^{3/2}(1/z^4)(\rm{cm}^{-3}/N_i)~~s	 \\ 
\tau_{i-e} & \sim & [1.4\times10^9](m_i/m_p)^{1/2}		\nonumber\\
& & ((kT_e/I_H)+5.4\times10^{-4}(kT_i/I_H)(m_p/m_i))^{3/2}  \nonumber\\
& & (1/z^2)(\rm{cm}^{-3}/N_i)~~s	 				\nonumber\\ 
\tau_{e-e} & \sim & [1.6\times10^6](kT_e/I_H)^{3/2}(\rm{cm}^{-3}/N_i)~~s.	 		\nonumber 
\end{eqnarray}
The ion mass is $m_i$, the proton mass  $m_p$, the ionisation  potential $\chi$, the ion density  $N_i$,
the electron density  $N_e$, the ion temperature  $T_i$, the electron temperature $T_e$ and the ionisation energy of hydrogen is
$I_H$. $\tau_{red}$ may span across the inequalities of equation \ref{eqn:eqn6} and is discussed in a later paragraph.     

\begin{figure}[htp]
\begin{center}
   \begin{minipage}[t]{10.0cm}
   \raggedright
\ \psfig{file=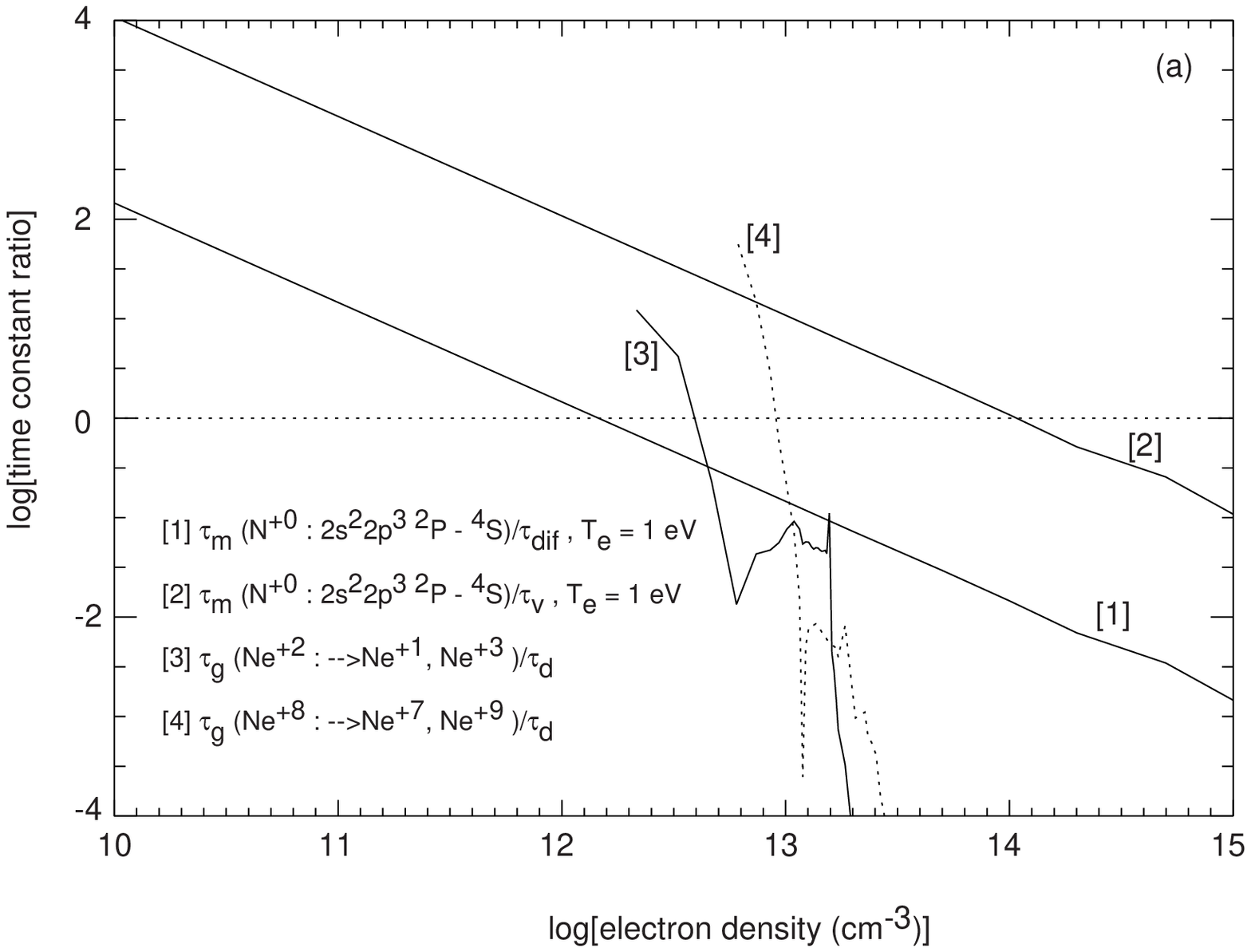,width=9.5cm}
\ \psfig{file=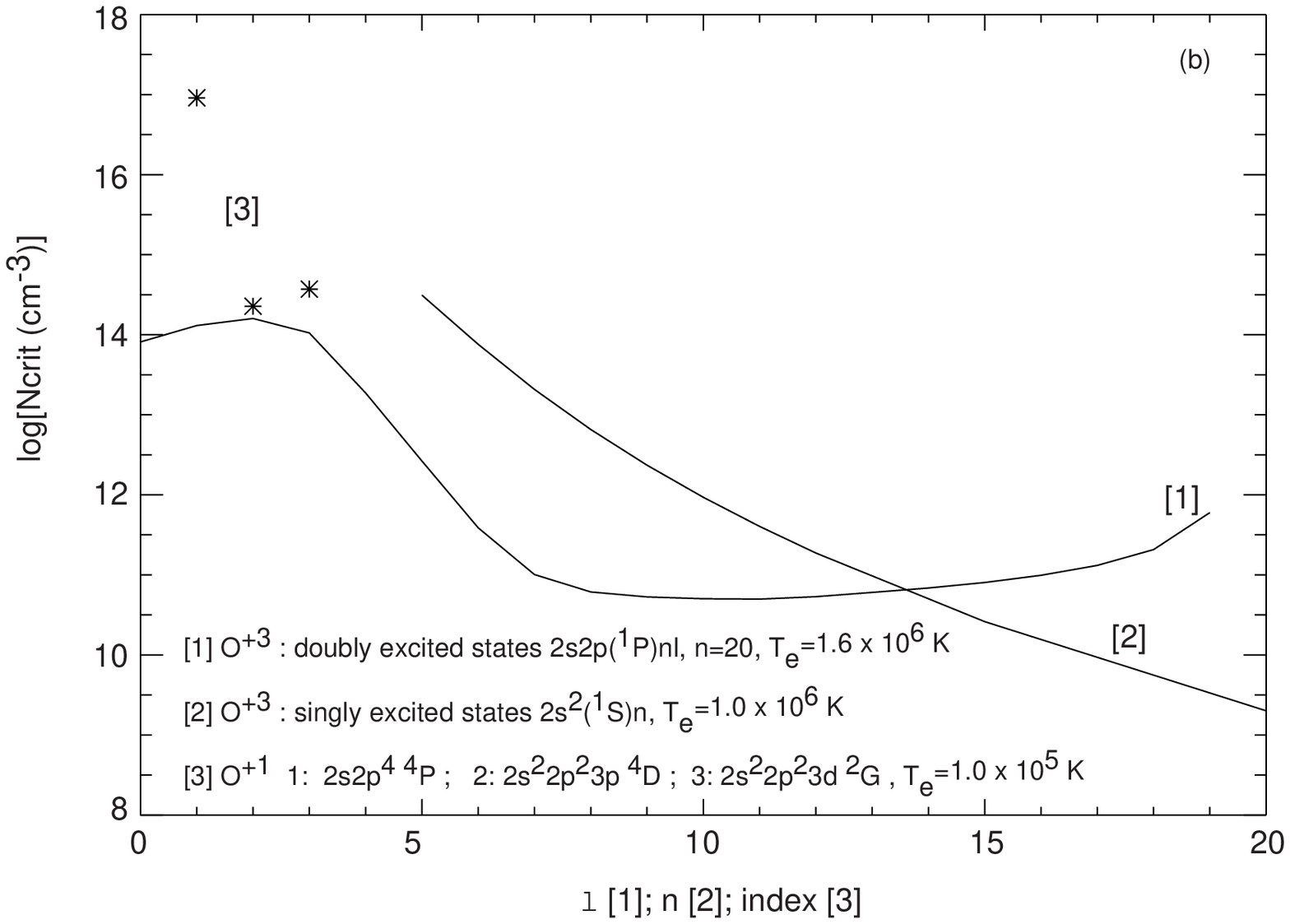,width=10.0cm}
   \end{minipage}
\caption{\label{fig:fig1} (a) Ratios of ground and metastable lifetimes to plasma timescales arising from transport of
some ions of nitrogen and neon at the edge of fusion plasmas.  For $\rm{N}^0$  the ground state to metastable state
transfer time constant, $\tau_m$ is contrasted with $\tau_v=\lambda_{sol}/v$ with typical scrape-off-layer thickness
$\lambda_{sol} \sim 2~\rm{cm}$ and $mv^2/2 \lesssim 1~\rm{eV}$ from chemical or physical sputtering and with
$\tau_{dif}=\lambda_{sol}^2/D$ where a typical diffusion coefficient $D=10^4~{\rm cm}^2 {\rm s}^{-1}$.  For selected
neon ions, $\tau_d=\lambda_{T_e,N_e}/v_{dif}$ is contrasted with the reciprocal of the sum of the inverses of the
ionisation and recombination time constants for the ground state. The latter illustrative results are for a JET-like
tokamak H-mode radial plasma model with shaped diffusion and pinch terms. $T_e(r=0)$=1 keV with a pedestal of 30eV at
$r=a$ and exponential decay (scale length $= 0.01a$) in the scrape of layer at plasma minor radius $a= 100~\rm{cm}$.
$v_{dif}$ combines pinch and concentration diffusion parts. (b) Critical densities, defined as $N_e(\tau_{red} =
\tau_{0,a})$, for categories of excited populations of some oxygen ions. The x-axis scale for curve [1] is the orbital
angular momentum $l$ of the outer electron of doubly excited states; for curve [2] it is the principal quantum number
$n$ of the singly excited electron; for curve [3] it is a simple index to the three low-lying states illustrated.  For low
levels of ions, $\tau_{red}$ is markedly sensitive to the detailed atomic structure.  $n=3$ valence shells populations
are of special relevance to light element spectroscopy in the visible.}
\end{center}
\end{figure}

From a dynamic point of view, the intrinsic and extrinsic groups are to be compared with each other and with timescales,
$\tau_{plasma}$, representing plasma ion diffusion across temperature or density scale lengths, relaxation times of
transient phenomena and observation times.  For most plasmas in magnetic confinement fusion and astrophysics
\begin{eqnarray}
\label{eqn:eqn8}
	\tau_{plasma} \sim \tau_g \sim \tau_m >> \tau_o >> \tau_{e-e}
\end{eqnarray}
where $\tau_g$ represents the relaxation time of ground state populations of ions (a composite of $\tau_{rec}$ and $\tau_{ion}$) and it is such
plasmas which are addressed in this paper.  These time-scales imply that the dominant populations of impurities in the plasma are  those of the
ground and metastable states of the various ions.  The dominant populations  evolve on time-scales of the order of plasma diffusion time-scales
and so should be modelled dynamically, that is in the time-dependent, spatially varying, particle number continuity equations, along with the
momentum and energy equations of plasma transport theory.  Illustrative results are shown in figure \ref{fig:fig1}a.

The excited populations of impurities and the free electrons on the other hand may be assumed relaxed with respect to the
instantaneous dominant populations, that is they are in a {\it quasi-equilibrium}.  The quasi-equilibrium is determined by
local  conditions of electron temperature and electron density.  So, the atomic modelling may be  partially de-coupled from
the impurity transport problem into local calculations which  provide quasi-equilibrium excited ion populations and
emissivities and then effective source coefficients (collisional-radiative coefficients) for dominant populations which must
be entered into the plasma transport equations.  The solution of the transport equations establishes the  spatial and
temporal behaviour of the dominant populations which may then be re-associated  with the local emissivity calculations for
matching to and analysis of observations.  

For excited populations, $\tau_{red}$ plays a special and complicated role due to the very large variation in collisional
excitation/de-excitation reaction rates with the quantum  numbers of the participating states.  In the low density coronal
picture $\tau_{red} >> \tau_o$ and redistribution plays no part.  Critical densities occur for  $\tau_{red} \sim \tau_o$ and
for $\tau_{red} \sim \tau_a$ and allow division of the (in principle) infinite number of excited populations into categories
including {\it low levels}, {\it high singly excited levels} and {\it doubly excited levels} for which important
simplifications are possible.  These are examined in section \ref{sec:sec2}.   Light element ions in fusion plasmas are
generally in the singly excited state redistibutive case, approaching the doubly excited redistribution case at the higher
densities.  Highly ionised ions of heavy species in fusion plasmas approach the coronal picture.  Illustrative results on
critical densities are shown in figure \ref{fig:fig1}b.

Finally, because of the generally short $\tau_{e-e}$ compared with other timescales (including those of free-free
and free-bound emission), it is usually the case that the free electrons have close to a Maxwellian distribution.  This
assumption is made throughout the present paper, but is relaxed in the next paper of the series (Bryans \etal, 2005).            
       
\subsection{Generalised collisional-radiative theory}
\label{sec:sec1.2}

The basic model was established by Bates \etal (1962). The ion in a plasma is viewed as  composed of a complete set of levels
indexed by $i$ and $j$ and a set of radiative and  collisional couplings between them denoted by $C_{ij}$ (an element of the
{\it collisional-radiative  matrix} representing transition from $j$ to $i$) to which are added direct ionisations from each  level of
the ion to the next ionisation stage (coefficient  $S_i$) and direct recombinations to each  level of the ion from the next
ionisation stage (coefficient $r_i$). Thus, for each level, there is a total loss rate coefficient for its population number density,
$N_i$, given by
\begin{eqnarray}
\label{eqn:eqn9}
 -C_{ii} = \sum_{j \ne i}C_{ji}~+N_eS_i.	
\end{eqnarray}
Following the discussion in the introduction, it is noted that populated metastable states can exist and  there is no
real  distinction between them and ground states.  We use the term {\it metastables} to denote both ground and
metastables states.  Metastables are the dominant populations and so only recombination events which start with a
metastable as a collision partner matter.  We condider the population structure of the $z$-times ionised ion, called
the recombined or child ion.  The $(z+1)$-times ionised ion is called the recombining or parent ion and the
$(z-1)$-times ionised ion is called the grandchild.  The metastables of the recombined ion are indexed  by $\rho$ and
$\sigma$, those of the recombining ion by $\nu$ and $\nu^{\prime}$  and those of the grandchild by $\mu$ and
$\mu^{\prime}$.  Therefore the ion of charge state $z$ has metastable populations  $N_{\rho}$, the recombining ion of
charge $(z+1)$ has metastable populations $N^+_{\nu}$ and the grandchild ion of charge $(z-1)$ has metastable
populations $N^-_{\mu}$.  We designate the remaining excited states of the $z$-times ionised ion, with the metastables
separated, as {\it ordinary} levels for which we reserve the indices $i$ and $j$ and populations $N_i$ and $N_j$.
There are then, for example, direct recombination coefficients $r_{i,\nu}$  from each parent metastable into each
child ordinary level and direct ionisation coefficients from each child ordinary level to each parent metastable
$S_{\nu,i}$ such that $S_i= \sum_{\nu}S_{\nu,i}$. Also there are direct ionisation coefficients $S_{\rho,\mu^{\prime}}
$to the metastables of the child from the metastables of the grandchild.  Then the continuity equations for population
number densities are   
\begin{eqnarray}
\label{eqn:eqn10}
	\frac{d}{dt}\left[\begin{array}{l}
		N^-_{\mu}  		\\
		N_{\rho}  		\\
		N_i			\\
		N^+_{\nu}  
	       \end{array}
	 \right]
	 =
	\left[\begin{array}{llll}
		\scriptnew{C}_{\mu \mu'} & N_e~\scriptnew{R}_{\mu,\sigma} & 0 & 0 				\\
		N_e\scriptnew{S}_{\rho \mu'} & C_{\rho \sigma} & C_{\rho j} & N_e~r_{\rho \nu'}	\\
		0  & C_{i \sigma} & C_{i j} & N_e~r_{i \nu'}		\\
		0 & N_eS_{\nu \sigma} & N_eS_{\nu j} & \scriptnew{C}_{\nu \nu'} 
	      \end{array}
	 \right]
	\left[\begin{array}{l}
		N^-_{\mu'}  		\\
		N_{\sigma}  		\\
		N_{j}  			\\
		N^+_{\nu'}  
	       \end{array}
	 \right]		\nonumber \\
\end{eqnarray}
where the equations for the $(z-1)$-times and $(z+1)$-times ionised ions have been simplified by incorporating their
ordinary population contributions in their metastable contributions (shown as script capital symbols) as the immediate focus
is on the $z$-times ionised ion.  This incorporation procedure is shown explicitly in the following equations for the
$z$-times ionised ion through to equations \ref{eqn:eqn16} and may be done for each ionisation stage separately.  Note
additionally the assumption (made by omission of the (3,1) partition element, where 3 denotes the row and 1 the
column) that state-selective ionisation from the stage $(z-1)$ takes place only into the metastable manifold of the
stage $z$.

\subsubsection{Derived source term coefficients}
\label{sec:sec1.2.1}
From the quasi-static assumption, we set $dN_i/dt = 0$ and then the matrix equation for the ordinary
levels of the $z$-times ionised ion gives
\begin{eqnarray}
\label{eqn:eqn11}
N_j & = & -C^{-1}_{ji}C_{i \sigma}N_{\sigma}-N_e C^{-1}_{j i}r_{i \nu'}N^+_{\nu'}	
\end{eqnarray}
where we have used summation convention on repeated indices.  Substitution in equations \ref{eqn:eqn10} for the metastables of the
$z$-times ionised ion gives 
\begin{eqnarray}
\label{eqn:eqn12}
\frac{dN_{\rho}}{dt} & = & N_e\left[ S_{\rho \mu'}\right]N^-_{\mu'}    \nonumber \\
 & & +\left[ C_{\rho \sigma}-C_{\rho j}C^{-1}_{ji}C_{i \sigma}\right] N_{\sigma} \nonumber \\
 & & + N_e\left[ r_{\rho \nu'}- C_{\rho j}C^{-1}_{ji}r_{i \nu'}\right]N^+_{\nu'}. 	
\end{eqnarray}
The left-hand-side is interpreted as a total derivative with time-dependent and convective parts and the right-hand-side comprises the
source terms. The terms in square brackets in equations \ref{eqn:eqn12} give the effective growth rates of each metastable population of the $z$-times ionised
ion driven by excitation (or de-excitation) from other metastables of the $z$-times ionised ion, by ionisation to the $(z+1)$-times ionised ion and excitation
to other metastables of the $z$-times ionised ion (a negatively signed growth) and by recombination from the metastables of the $(z+1)$-times ionised ion. 
These are called the $\scriptnew{GCR}$ coefficients.  Following Burgess and Summers (1969), who used the name `collisional-dielectronic' for
`collisional-radiative' when dielectronic recombination is active, we use the nomenclature $\scriptnew{ACD}$ for the $\scriptnew{GCR}$ recombination
coefficients which become   
\begin{equation}
\label{eqn:eqn13}
\scriptnew{ACD}_{\nu \rightarrow \rho} \equiv \scriptnew{R}_{\rho \nu} =  r_{\rho \nu}- C_{\rho j}C^{-1}_{ji}r_{i \nu}. 	
\end{equation}
The $\scriptnew{GCR}$ metastable cross-coupling coefficients (for $\rho \ne \sigma$) are 
\begin{equation}
\label{eqn:eqn14}
\scriptnew{QCD}_{\sigma \rightarrow \rho} \equiv \scriptnew{C}_{\rho \sigma}/N_e =  \left[ C_{\rho \sigma}-C_{\rho j}C^{-1}_{ji}C_{i
\sigma}\right] /N_e. 	
\end{equation}
Note that the on-diagonal element $\left[ C_{\rho \sigma}-C_{\rho j}C^{-1}_{ji}C_{i \sigma}\right] /N_e$ with $\sigma=\rho$ is a
total loss rate coefficient from the metastable $\rho$.  Substitution from equation \ref{eqn:eqn11} in the equations
\ref{eqn:eqn10} for the metastables of the $z+1$-times ionised ion gives 
\begin{eqnarray}
\label{eqn:eqn15}
\frac{dN^+_{\nu}}{dt} & = & N_e\left[ S_{\nu \sigma}-S_{\nu j}C^{-1}_{ji}C_{i \sigma}\right] N_{\sigma} \nonumber \\
 & & +\left[ C_{\nu \nu'}- N_e^2S_{\nu j}C^{-1}_{ji}r_{i \nu'}\right]N^+_{\nu'}. 	
\end{eqnarray}
The  $\scriptnew{GCR}$ ionisation coefficients resolved by initial and final metastable state are   
\begin{equation}
\label{eqn:eqn16}
\scriptnew{SCD}_{\sigma \rightarrow \nu} \equiv \scriptnew{S}_{\nu \sigma}  = \left[ S_{\nu \sigma}-S_{\nu j}C^{-1}_{ji}C_{i \sigma}\right]  	
\end{equation}
and note that there is contribution to cross-coupling between parents via recombination to excited states of the $z$-times ionised
ion followed by re-ionisation to a different metastable,
\begin{equation}
\label{eqn:eqn17}
\scriptnew{XCD}_{\nu' \rightarrow \nu}  = -N_e\left[ S_{\nu j}C^{-1}_{ji}r_{i \nu'}\right]. 	
\end{equation}

Consider the sub-matrix comprising the (2,2), (2,3), (3,2) and (3,3) partitions of equations \ref{eqn:eqn10}.  Introduce the inverse
of this sub-matrix as
\begin{eqnarray}
\label{eqn:eqn18}
	\left[\begin{array}{ll}
	        W_{\rho \sigma} & W_{\rho j}                  \\
		W_{i \sigma}    & W_{i j}                     \\
	      \end{array}
	 \right]
	 =
	\left[\begin{array}{ll}
	        C_{\rho \sigma} & C_{\rho j}                  \\
		C_{i \sigma}    & C_{i j}                     \\
	      \end{array}
	 \right]^{-1}
\end{eqnarray}
and note that the inverse of the (1,1) partition
\begin{equation}
\label{eqn:eqn19}
\left[W_{\rho \sigma}\right]^{-1} \equiv \scriptnew{C}_{\rho \sigma}= \left[
C_{\rho \sigma}-C_{\rho j}C^{-1}_{ji}C_{i \sigma}\right].
\end{equation}
This compact representation illustrates, that the imposition of the quasi-static assumption leading to elimination of the
ordinary level populations in favour of the metastable populations, may be viewed as a {\it condensation} in which the
influence of the ordinary levels is {\it projected} onto the metastable levels.  The metastables can be condensed in a
similar manner onto the ground restoring the original (ground states only) collisional-radiative picture.  The additive
character of the direct metastable couplings $C_{\rho \sigma}$ means that these elements may be adjusted retrospectively
after the main condensations.

\subsubsection{Derived emission and power coefficients}
\label{sec:sec1.2.2}
There are two kinds of derived coefficients associated with individual spectrum line emission 
in common use in fusion plasma diagnosis.  These are {\it photon emissivity coefficients} (${\cal PEC}$) 
and  {\it ionisation per photon ratio} (${\cal SXB}$).  The reciprocals of the latter are also known as 
{\it photon efficiencies}.  
From equations \ref{eqn:eqn11}, the emissivity in the spectrum line $j \rightarrow k$ may be written as
\begin{eqnarray}
\label{eqn:eqn20}
	\epsilon_{j \rightarrow k} & = & =  A_{j \rightarrow k}N_eN_j = A_{j \rightarrow k}(\sum_{\sigma}{\cal F}^{(exc)}_{j \sigma}N_eN_{\sigma} +
	        \sum_{\nu'=1}^M{\cal F}^{(rec)}_{j \nu'}N_eN^+_{\nu'}).  				     
\end{eqnarray}
This allows specification of the {\it excitation} photon emissivity coefficient
\begin{eqnarray}
\label{eqn:eqn21}
	\scriptnew{PEC}^{(exc)}_{\sigma,j \rightarrow k} = A_{j \rightarrow k}{\cal F}^{(exc)}_{j \sigma}
\end{eqnarray}
and the {\it recombination} photon emissivity coefficient
\begin{eqnarray}
\label{eqn:eqn22}
	\scriptnew{PEC}^{(rec)}_{\nu',j \rightarrow k} = A_{j \rightarrow k}{\cal F}^{(rec)}_{j \nu'}.
\end{eqnarray}
The ionisation per photon ratios are most meaningful for the excitation part of the emissivity and are
\begin{eqnarray}
\label{eqn:eqn23}
	\scriptnew{SXB}^{(ion)}_{\sigma,j \rightarrow k} = \sum_{\nu=1}^{M_z-1}\scriptnew{SCD}_{\sigma \rightarrow \nu}/
	A_{j \rightarrow k}{\cal F}^{(exc)}_{j \sigma}.
\end{eqnarray}
Each of these coefficients is associated with a particular metastable $\sigma$, $\nu'$ or  $\mu'$ of the 
$A^{+z}$, $A^{+z+1}$ or $A^{+z-1}$ ions respectively.

The radiated power in a similar manner separates into parts driven by excitation and by recombination as 
\begin{eqnarray}
\label{eqn:eqn24}
	\scriptnew{PLT}_{\sigma} = \sum_{j,k}\Delta{E}_{j \rightarrow k}A_{j \rightarrow k}{\cal F}^{(exc)}_{j \sigma}
\end{eqnarray}
called the {\it low-level line power coefficient} and 
\begin{eqnarray}
\label{eqn:eqn25}
	\scriptnew{PRB}_{\nu'} = \sum_{j,k}\Delta{E}_{j \rightarrow k}A_{j \rightarrow k}{\cal F}^{(rec)}_{j \nu'}
\end{eqnarray}
called the {\it recombination-bremsstrahlung power coefficient} where it is convenient to include bremsstrahlung with $\scriptnew{PRB}$.  Note that in the
generalised picture, additional power for the $z$-times ionised ions occurs in forbidden transitions between metastables as
\begin{eqnarray}
\label{eqn:eqn26}
\sum_{\rho,\sigma}\Delta{E}_{\sigma \rightarrow \rho }A_{\sigma \rightarrow \rho }N_{\sigma}
\end{eqnarray}
for the $z$-times ionised ion.  In the fusion context, this is usually small.  Radiated power is the most relevant
quantity for experimental detection. For modelling, it is the {\it electron energy loss function} which enters the fluid energy
equation.  The contribution to the total electron energy loss rate for the $z$-times ionised ion associated with ionisation and recombination
from the $(z+1)$-times ionised ion is
\begin{eqnarray}
\label{eqn:eqn27}
	&&\sum_{\rho}E_{\rho} (\sum_{\sigma}\scriptnew{C}_{\rho \sigma}N_{\sigma}+Ne\sum_{\nu'}\scriptnew{R}_{\rho
          \nu'}N_{\nu'}) +  \nonumber \\
	&&\sum_{\nu}E_{\nu} (\sum_{\sigma}\scriptnew{S}_{\nu \sigma}N_{\sigma}+Ne\sum_{\nu'}\scriptnew{C}_{\nu
          \nu'}N_{\nu'}) +  \nonumber \\
	&&\sum_{\rho,\sigma}\Delta{E}_{\sigma \rightarrow \rho }A_{\sigma \rightarrow \rho }N_{\sigma}
	+\sum_{\sigma}\scriptnew{PLT}_{\sigma}N_eN_{\sigma}+ \nonumber \\
	&&\sum_{\nu'}\scriptnew{PRB}_{\nu'}N_eN_{\nu'} 
\end{eqnarray}
where the $E_{\rho}$ and $E_{\nu}$ are absolute energies of the metastables $\rho$ of the $z$-times ionised ion and of the metastables $\nu$ of the $(z+1)$-times
ionised ion respectively. Thus the electron energy loss is a derived quantity from the radiative power coefficients and the other generalised collisonal-radiative
coefficients.  Note that cancellations in the summations cause the reduction to relative energies and that in ionisation equilibrium, the electron energy loss
equals the radiative power loss.   

In the following sections, it is shown how the quasi-static assumption and population categorisation allow us to solve the
infinite level population structure of each ion in a manageable and efficient way.  Such solution is necessary  for low and
medium density astrophysical and magnetic confinement fusion plasmas. This is unlike the simpler situation of very dense
plasmas where heavy level truncation is used, because of continuum merging.

\section{Excited population structure}
\label{sec:sec2}

The handling of metastables in a generalised collisonal-radiative framework requires a detailed specific
classification of level structure compatible with both recombining and recombined ions.  For light element ions,
Russell-Saunders (L-S) coupling is appropriate and it is sufficient to consider only terms, since although fine
structure energy separations may be required for high resolution spectroscopy, relative populations of levels of a
term are close to statistical.  So the parent ion metastables are of the form  $\gamma_{\nu}L_{\nu}S_{\nu}$ with
$\gamma_{\nu}$ the configuration and the recombined ion metastables are of the form $\Gamma_{\rho}L_{\rho}S_{\rho}$
with the configuration $\Gamma_{\rho}= \gamma_{\nu} + n_il_i$ and the excited (including highly excited) terms are
$(\gamma_{\nu}L_{\nu}S_{\nu})n_il_i L_i S_i$.  $n$ and $l$ denote individual electron principal quantum number and
orbital angular momentum, $L$ and $S$ denote total orbital and total spin angular momenta of the multi-electron ion
respectively in the specification of an ion state in Russell-Saunders coupling.  The configuration specifies the
orbital occupancies of the ion state.  For ions of heavy elements, relative populations of fine structure levels can
differ markedly from statistical and it is necessary to work in intermediate coupling with parent metastables of the
form  $\gamma_{\nu}J_{\nu}$ and recombined metastables of the form $\Gamma_{\rho}J_{\rho}$ with the configuration
$\Gamma_{\rho}= \gamma_{\nu} + n_il_i$ and the excited (including highly excited) levels
$(\gamma_{\nu}J_{\nu})n_il_ij_i J_i$.  There is a problem.  To cope with the very many principal quantum shells
participating in the calculations of collisional-dielectronic coefficients at finite density necessitates a grosser
viewpoint (in which populations are {\it bundled}), whereas for modelling detailed spectral line emission, the finer
viewpoint (in which populations are fully {\it resolved}) is required.   In practice, each ion tends to have a limited
set of low levels  principally responsible for the dominant spectrum line power emission for which a bundled approach
is too imprecise, that is, averaged energies, oscillator strengths and collision  strengths do not provide a good
representation.  Note also that key parent transitions for  dielectronic recombination span a few (generally the same)
low levels for which precise atomic data are necessary.   In the recombined ion, parentage gives approximate quantum
numbers, that is, levels of the  same $n$ (and $l$) divide into those based on different parents.  Lifetimes of levels
of the same $n$ but  different parents can vary strongly (for example through secondary autoionisation).  Also the 
recombination population of such levels is generally from the parent with which they are  classified. 
\begin{figure}[htp]
\begin{center}
   \begin{minipage}[t]{8.4cm}
   \raggedright
\ \psfig{file=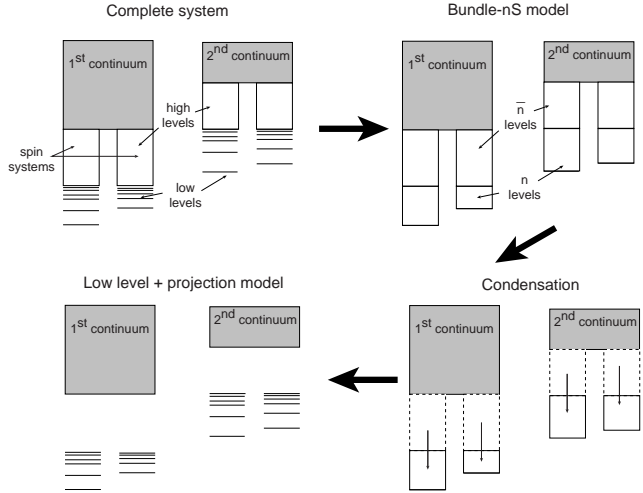,width=8.4cm}
   \end{minipage}
\caption{\label{fig:fig2} Schematic of population modelling and condensation procedures. It represents two metastable parent states of non-zero spin, so that when
an $nl$ electron is added, each yields two spin systems for the $z$-times ionised ion.}
\end{center}
\end{figure}
We therefore recognise three sets of non-exclusive levels of the recombined ion:
\begin{itemize} 
\item[(i)] Metastable levels - indexed by $\rho$, $\sigma$.
\item[(ii)] Low levels - indexed by $i$, $j$ in a resolved coupling scheme, being the complete set of 
levels of a principal quantum shell range $n:n_0 \le n \le n_1$, including relevant 
metastables and spanning transitions contributing substantially to radiative power or 
of interest for specific observations.
\item[(iii)] Bundled levels - segregated according to the parent metastable upon which they are 
built and possibly also by spin system - which can include {\it bundle-nl} and {\it bundle-n}.
\end{itemize}
Viewed as a recombining ion, the set (i) must include relevant parents and set (ii) must span  transitions which are dielectronic
parent transitions.  Time dependence matters only for the  populations of (i), high precision matters only for groups (i) and (ii)
and special very many  level handling techniques matter only for group (iii).  

To satisfy the various requirements and  to allow linking of population sets at different resolutions, a series of manipulations on
the  collisional-radiative matrices are performed (Summers and Hooper, 1983).  To illustrate this, suppose there is a single parent
metastable state.   Consider the collisional-radiative matrix for the recombined ion and the right hand side in the bundle-n
picture, and a partition of the populations as $[n,\bar{n}]$  with  $n:n_0 \le n \le n_1$  and  $\bar{n}:n_1 \le \bar{n}$. 
Elimination of the  $N_{\bar{n}}$ yields a set  of equations for the  $N_n$.  We call this a `condensation' of the whole set of
populations onto the $n$ populations.  The  coefficients are the effective ionisation coefficients from the $n$, the effective
cross-coupling  coefficients between the $n$ and the effective recombination coefficients into the $n$, which now  include direct
parts and indirect parts through the levels $\bar{n}$.  Exclusion of the direct terms prior  to the manipulations yields only the
the indirect parts.  Call these  $C^{indir}_{nn'}$ and  $r^{indir}_{n\nu}$.  We make  the assumption that  $C^{indir}_{nn'}$ and 
$r^{indir}_{n\nu}$ may be expanded over the resolved low level set (see section \ref{sec:sec2.2}) to give the expanded indirect
matrix  $C^{indir}_{ij}$ and  $r^{indir}_{i\nu}$ where  $i$ and $j$ span  the resolved low level set (ij).  These indirect
couplings are then combined with higher precision direct couplings  $C^{dir}_{ij}$  and  $r^{dir}_{i\nu}$ so that 

\begin{eqnarray}
\label{eqn:eqn28}
C_{ij}  =C^{dir}_{ij}+C^{indir}_{ij}
\end{eqnarray}
and
\begin{eqnarray}
\label{eqn:eqn29}
& r_{i\nu} & =r^{dir}_{i\nu}+r^{indir}_{i\nu}.	
\end{eqnarray}
The procedure is shown schematically in figure \ref{fig:fig2}.  The process may be continued, condensing the low level set onto the
metastable set.  The generalised collisional-dielectronic coefficients are the result.   The time dependent and/or spatial
non-equilibrium transport equations which describe the  evolution of the ground and metastable populations of ions in a plasma use
these generalised  coefficients.  Following solution, the condensations can be reversed to recover the complete  set of excited
populations and hence any required spectral emission.  The progressive  condensation described above can be viewed as simply one of
a number of possible paths  which might be preferred because of special physical conditions or observations.

For light element ions, four types of bundling and condensation are distinguished in this work:
\begin{itemize}
\item[(a)] Ground parent, spin summed bundle-n $\rightarrow$ lowest n-shell.
\item[(b)] Parent and spin separated bundle-n $\rightarrow$ lowest spin system n-shell (the {\it bundle-nS population model}).
\item[(c)] Low LS resolved $\rightarrow$ metastable states (the {\it low-level population model}).
\item[(d)] Parent and spin separated bundle-n $\rightarrow$ low LS resolved $\rightarrow$ metastable states.
\end{itemize}
Type (a) corresponds to the approach used in Summers (1974).  Type (c) corresponds to the  usual population calculation for low
levels in which (consistent) recombination and ionisation involving excited states are ignored.  It establishes the dependence of
each population on excitation for the various  metastables only.  Type (d), effectively the merging of (b) and (c) in the manner
described earlier, is the principal procedure to be  exploited in this work for first quality studies.   Details are in the
following sub-sections.

\subsection{The bundle-$nS$ model}
\label{sec:sec2.1}
\begin{table}
\begin{center}
\begin{tabular}{cccccc}
Rec.     & Parent & Parent/spin             &  $\omega_{S_{\nu},S}$  & Lowest              & $N_{met}$  \\
seq.     & index  & system                  &                        & metastable          &	        \\
\hline
H-like   &  1     & $(1s~^2S)~^1n$          & 0.250                  & $1s^2~^1S$          &   1        \\   
         &        & $(1s~^2S)~^3n$          & 0.750                  & $1s2s~^3S$          &   1        \\   
He-like  &  1     & $(1s^2~^1S)~^2n$        & 1.000                  & $1s^22s~^2S$        &   1        \\   
         &  2     & $(1s2s~^3S)~^2n$        & 0.333                  & $1s2s^2~^2S$        &   1        \\   
         &        & $(1s2s~^3S)~^4n$        & 0.667                  & $1s2s2p~^4P$        &   1        \\   
Li-like  &  1     & $(2s~^2S)~^1n$          & 0.250                  & $2s^2~^1S$          &   1        \\   
         &        & $(2s~^2S)~^3n$          & 0.750                  & $2s2p~^3P$          &   1        \\   
Be-like  &  1     & $(2s^2~^1S)~^2n$        & 1.000                  & $2s^22p~^2P$        &   1        \\   
         &  2     & $(2s2p~^3P)~^2n$        & 0.333                  & $2s^22p~^2P$        &   1        \\   
         &        & $(2s2p~^3P)~^4n$        & 0.667                  & $2s2p^2~^4P$        &   1        \\   
B-like   &  1     & $(2s^22p~^2P)~^3n$      & 0.750                  & $2s^22p^2~^1D$      &   2        \\   
         &        & $(2s^22p~^2P)~^1n$      & 0.250                  & $2s^22p^2~^3P$      &   1        \\   
         &  2     & $(2s2p^2~^4P)~^3n$      & 0.375                  & $2s^22p^2~^3P$      &   1        \\   
         &        & $(2s2p^2~^4P)~^5n$      & 0.625                  & $2s2p^3~^5S$        &   1        \\   
C-like   &  1     & $(2s^22p^2~^3P)~^4n$    & 0.667                  & $2s^22p^3~^4S$      &   1        \\   
         &        & $(2s^22p^2~^3P)~^2n$    & 0.333                  & $2s^22p^3~^2D$      &   2        \\   
         &  2     & $(2s^22p^2~^1D)~^2n$    & 1.000                  & $2s^22p^3~^2D$      &   2        \\   
         &  3     & $(2s^22p^2~^1S)~^2n$    & 1.000                  & $2s^22p^3~^2D$      &   2        \\   
         &  4     & $(2s2p^3~^5S)~^4n$      & 1.000                  & $2s^22p^3~^4S$      &   1        \\   
N-like   &  1     & $(2s^22p^3~^4S)~^3n$    & 0.375                  & $2s^22p^4~^3P$      &   1        \\   
         &        & $(2s^22p^3~^4S)~^5n$    & 0.625                  & $2s^22p^33s~^5S$    &   1        \\   
         &  2     & $(2s^22p^3~^2D)~^3n$    & 0.750                  & $2s^22p^4~^3P$      &   1        \\   
         &        & $(2s^22p^3~^2D)~^1n$    & 0.250                  & $2s^22p^4~^1D$      &   2        \\   
         &  3     & $(2s^22p^3~^2P)~^3n$    & 0.750                  & $2s^22p^4~^3P$      &   1        \\   
         &        & $(2s^22p^3~^2P)~^1n$    & 0.250                  & $2s^22p^4~^1D$      &   2        \\   
O-like   &  1     & $(2s^22p^4~^3P)~^2n$     & 0.250                  & $2s^22p^5~^2P$      &   1        \\   
         &        & $(2s^22p^4~^3P)~^4n$     & 0.250                  & $2s^22p^43s~^4P$    &   1        \\   
         &  2     & $(2s^22p^4~^1D)~^2n$     & 1.000                  & $2s^22p^43s~^2D$    &   2        \\   
         &  3     & $(2s^22p^4~^1S)~^2n$     & 1.000                  & $2s^22p^43s~^2D$    &   2        \\   
F-like   &  1     & $(2s^22p^5~^2P)~^1n$     & 0.250                  & $2s^22p^6~^1S$      &   1        \\   
         &        & $(2s^22p^5~^2P)~^3n$     & 0.750                  & $2s^22p^53s~^3P$    &   1        \\   
         
\end{tabular}
\caption{\label{table:tab1}Bundle-$nS$ calculation pathways.  The parent/spin system weight factor is defined in equation
\ref{eqn:eqn32} and \ref{eqn:eqn33}. $N_{met}$ indicates the number of metastables of the recombined ion associated with the parent/spin system.  The {\it parent
index}, sequentially numbering the different parents shown in brackets, is used as the reference for tabulation of coefficients. }
\end{center}
\end{table}
Now let $\scriptnew{A}^{+z_1}$  denote the recombining ion and $\scriptnew{A}^{+z_1-1}$  the recombined ion so that $z=z_1-1$   is
the ion charge of the latter.  $z_1$  is the effective ion charge and takes the place of the nuclear  charge in the reduction of
hydrogenic rate coefficients to compact forms in the statistical  balance equations.  Also introduce bundled populations 
\begin{equation}
\label{eqn:eqn30}
N_{\nu,nS} \equiv N_{(\gamma_{\nu}L_{\nu}S_{\nu}),nS} =\sum _{l,L}N_{(\gamma_{\nu}L_{\nu}S_{\nu}),nlLS}
\end{equation}
and the assumption that 
\begin{equation}
\label{eqn:eqn31}
N_{\nu,nlLS}=\frac{(2L+1)}{(2l+1)(2L_{\nu}+1)}N_{\nu,nlS}=\frac{(2L+1)}{n^2(2L_{\nu}+1)}N_{\nu,nS}.
\end{equation}
The bundling is based on the observation that the largest collision cross-sections are those for  which $n=n'$ and $l=l'\pm 1$. 
For these cases the transition energy is small (effectively zero for hydrogen or hydrogenic ions) and the cross-sections are so
large for electron densities of relevance for fusion that it is very good  approximation to assume relative statistical population
for the $lL$ sub-states.  The assumption is weakest for populations of states with core penetrating valence electron orbitals and
we expect spin system breakdown for high $nl$ states progressively at lower $n$ and $l$  for increasing ion charge $z$.  Thus the
above assumptions are appropriate for light element ions with a more elaborate {\it bundle-}$(J_p)nlj$ model more suited to heavy
element ions.  The latter will be the subject of a separate paper. In the bundle-$nS$ model, only equilibrium populations of
complete n-shells for a given parent and spin system need be evaluated, which are the solutions of the statistical balance equations
\begin{eqnarray}
\label{eqn:eqn32}
& \sum_{n'=n+1}^{\infty}[A_{n' \rightarrow n}+N_eq^{(e)}_{n' \rightarrow n} + N_pq^{(p)}_{n'\rightarrow n}]N_{\nu,n'S} \nonumber \\
+ & \sum_{n''=n_0}^{n-1}[N_eq^{(e)}_{n'' \rightarrow n}+N_pq^{(p)}_{n''\rightarrow n}]N_{\nu,n''S}	\nonumber \\
+ & \omega_{S_{\nu},S} \left [ N_eN_{\nu}^+\alpha^{(r)}_n+N_eN_{\nu}^+\alpha^{(d)}_n+N^2_eN_{\nu}^+\alpha^{(3)}_n \right ] \nonumber\\
= & \{ \sum_{n'=n+1}^{\infty}[N_eq^{(e)}_{n \rightarrow n'}+N_pq^{(p)}_{n \rightarrow n''}]	 \\
+ & \sum_{n''=n_0}^{n-1}[A_{n \rightarrow n''}+N_eq^{(e)}_{n \rightarrow n''} + N_pq^{(p)}_{n \rightarrow n''}] \nonumber \\
+ & N_eq^{(e)}_{n \rightarrow \epsilon}+ N_eq^{(p)}_{n \rightarrow \epsilon} + \sum_{\nu'}A^a_{\nu,nS \rightarrow \nu'}\} N_{\nu,nS}. \nonumber
\end{eqnarray}
$N_{\nu}^+$ is the population of the parent ion $\scriptnew{A}^{+z_1}_{\nu}$, $N_e$  is the free  electron density and  $N_p$ the free proton density, $A$ is the
usual Einstein  coefficient,  $q^{(e)}$ and $q^{(p)}$  denote collisional rate coefficients due to electrons and protons, $\alpha^{(r)}_n$, $\alpha^{(d)}_n$  and 
$\alpha^{(3)}_n$ denote radiative, dielectronic and three-body recombination coefficients respectively, $A^a_{\nu,nS \rightarrow \nu'}$ denotes secondary
autoionisation and $q^{(e)}_{n \rightarrow \epsilon}$ and $q^{(p)}_{n \rightarrow \epsilon}$ denote collisional ionisation rate coefficients due to electrons and
protons respectively. $\omega_{S_{\nu},S}$ is a spin weight factor (see equation \ref{eqn:eqn33} below). For complex ions, there are separate systems of equations
for each parent and for up to two spin systems (in L-S coupling) built on each  parent and one such equation for each value of $n$ from the lowest allowed $n$-shell
$n_0$ for the parent/spin system to $\infty$.  The number of spin systems is labelled $N_{sys}$ and the lowest allowed n-shell $n_0=n_0^{(r)}$, the lowest
accessible shell by recombination except for doublets built on the He-like $1s2s~^3S$ parent.  In general bare nuclei of other elements are effective ion
projectiles along with protons. We use the word protons here to represent mean $z_{eff}$ ions with suitably scaled collisional rate coefficients.   These equations
are analogous to the equations for hydrogen and coupling-independent expressions may be used for the main $n \rightarrow n'$ coefficients providing a suitable spin
system weight factor 
\begin{equation}
\label{eqn:eqn33}
\omega_{S_{\nu},S}=\frac{(2S+1)}{2(2S_{\nu}+1)}
\end{equation}  
is introduced. 
Table \ref{table:tab1} summarises these various parameters for first period iso-electronic sequences up to fluorine/neon. There are a
number of issues.  

\subsubsection{b-factors and lowest levels}
\label{sec:sec2.1.1}
For hydrogenic ions it was advantageous to write  the statistical equations in terms of Saha-Boltzmann deviation
factors. This remains true for  complex ions but the definition must be generalised.   The deviation  $b_{\nu,nS}$ is defined by
\begin{eqnarray}
\label{eqn:eqn34}
N_{\nu,nS}=N_eN^+_{\nu}8 \left (\frac{\pi a_0^2I_H}{kT_e} \right )^{3/2}
\frac{\omega_{\nu,nS}}{2\omega_{\nu}}{\rm exp}(I_{\nu,nS}/kT_e)b_{\nu,nS}.
\end{eqnarray}
That is $b_{\nu,nS}$ is specified with respect to the parent ion state $\scriptnew{A}^{+z_1}_{\nu}$, and the ionisation potential $I_{\nu,nS}$  is also 
referred to that parent.  $a_0$ is the Bohr radius.  Note that $\omega_{\nu,nS}=\omega_{S_{\nu},S}n^2 \omega_{\nu}$ where
$\omega_{\nu}=(2S_{\nu}+1)(2L_{\nu}+1)$ is the parent statistical weight.    It is convenient to introduce  $c_{\nu,nS}=b_{\nu,nS}-1$
and scaled  temperatures and densities
\begin{eqnarray}
\label{eqn:eqn35}
\theta_e & = & \left ( \frac{kT_e}{I_H} \right ) \frac{1}{z^2_1}	\nonumber \\
\rho_e & = & 2^5\sqrt{\frac{\pi}{3}}\frac{\pi a^3_0}{\alpha^3}\frac{N_e}{z^7_1}
\end{eqnarray}
with similar forms for the proton temperature and density.  $\alpha$ is the fine structure constant.  In these terms the statistical  balance equations become
particularly suitable for calculation and resulting $\scriptnew{GCR}$ coefficients (z-scaled), on a $\theta_e/\rho_e$ grid, can sustain interpolation of moderate
precision within an iso-electronic sequence.

The parent/spin system model does not distinguish recombined  metastables within the same spin system. For example, consider the
$2s^22p~^2P$ parent in a B-like  like ion recombining into the singlet system of the C-like ion. The parent/spin system has  two
metastables, $2s^22p^2~^1D$ and $2s^22p^2~^1S$. We assign the effective ground state as the lowest energy metastable ($2s^22p^2~^1D$
state in this example) and assign statistically weighted n-shell sub-populations to resolve between the recombined metastables. There
is often a substantial difference between the  quantum defect of the lowest state itself and the mean quantum defect arising from a
weighted  average of the lowest n-shell terms. These choices improve the stepwise part of the collisional-radiative ionisation
coefficient within the bundle-$nS$ picture at low density and allow the bundle-$nS$ model to stand alone, albeit at some reduction in
accuracy.  A precise solution, as adopted in this paper, is given by the condensation/projection/expansion matrix transfer of the
bundle-$nS$ model  onto the low level solution which distinguishes explicitly between the LS terms in the lowest few  quantum shells
from the beginning (see section \ref{sec:sec2.2}).
\begin{figure}[htp]
\begin{center}
   \begin{minipage}[t]{8.4cm}
   \raggedright
\ \psfig{file=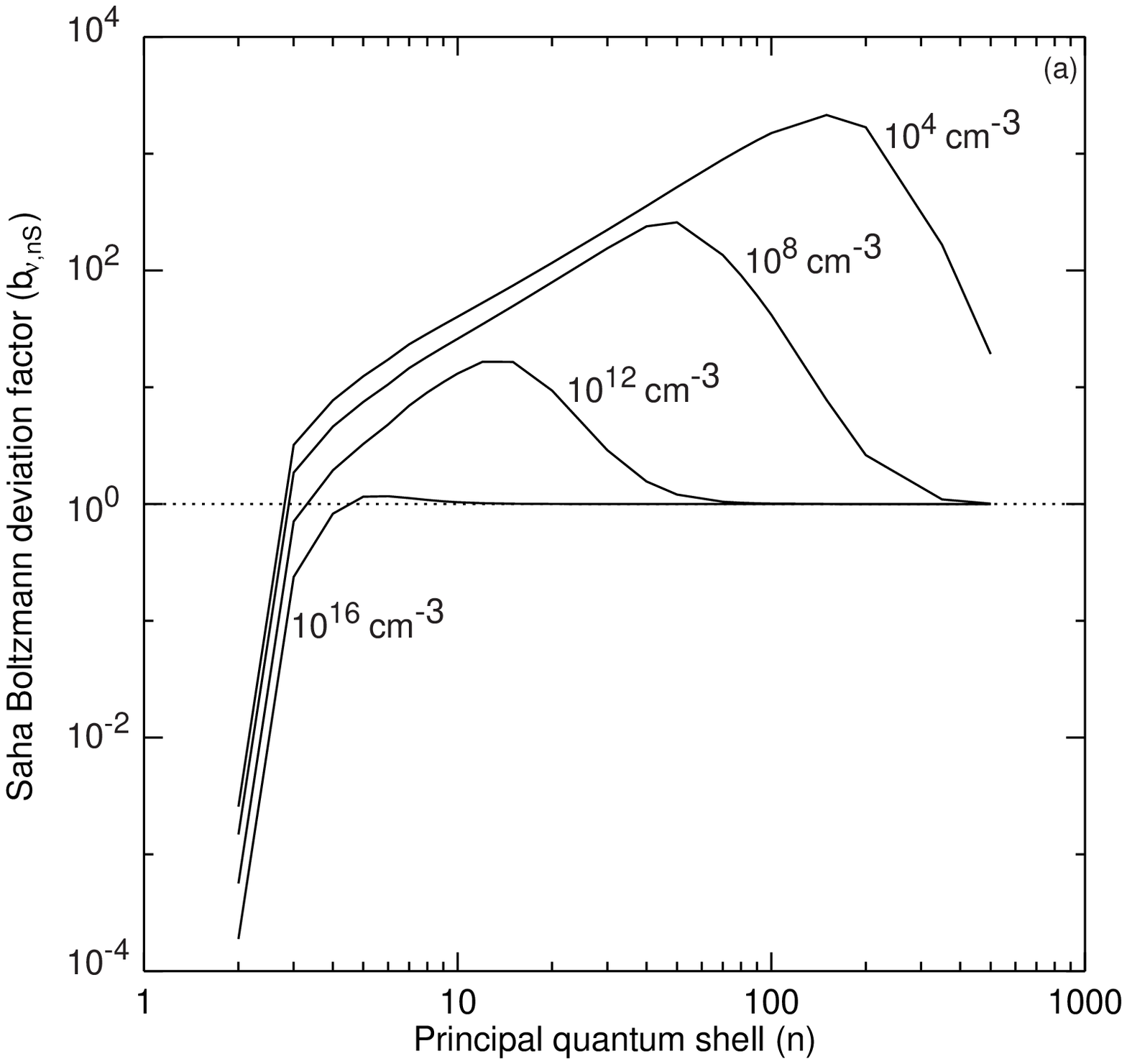,width=8.4cm}
\ \psfig{file=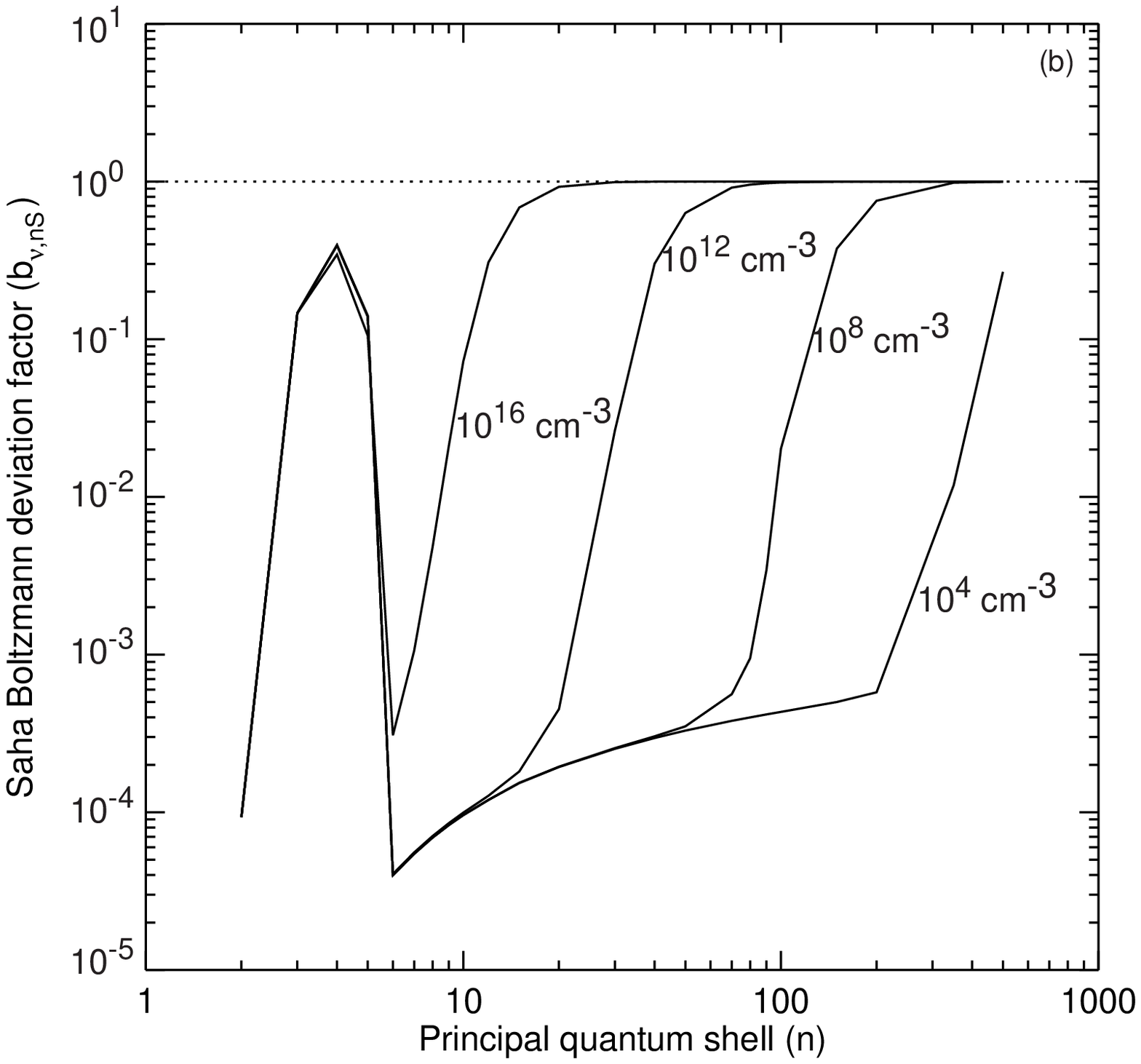,width=8.4cm}
   \end{minipage}
\caption{\label{fig:fig3} Bundle-$nS$ population structure for $\rm{O}^{+4}$ recombining from and ionising to $\rm{O}^{+5}$ at
$T_e=1.7 \times 10^5~\rm{K}$. (a) $b_{(2s^2~^1S)~^2n}$ factors for doublet bundle-n populations built on the ground parent $2s^2~^1S$ term.  The
enhanced b-factors at $n \sim 10-300$ is due to dielectronic recombination.  Curves show the progressive suppression of the high
populations as the electron density increases. (b) $b_{(2ssp~^3P)~^2n}$ factors for doublet bundle-n populations built
on the metastable parent $2s2p~^3P$ term.  The abrupt transition to underpopulations at $n \sim 4$ is due to secondary Auger
transition ($LS$ coupling allowed breakdown type) to the $^1S$ parent system.}
\end{center}
\end{figure}

\subsubsection{Auto-ionisation and alternate parents}
\label{sec:sec2.1.2}

Let the metastable state which represents the lowest level of recombined parent/spin system $\nu,nS$  be labelled by $\rho$. With
parent $\nu$, from a stepwise collisional-radiative point of view it  is convenient to refer to  $\nu,nS$ as an intermediate state
system and label it $(\nu)~^{(2S+1)}n$.  Thus a recombination/stepwise/cascade pathway may progress from a parent $\nu$ through the
intermediate state $\nu,nS$ ending on the metastable $\rho$ with the effective $\scriptnew{GCR}$ recombination coefficient
written as $\scriptnew{ACD}_{\nu \rightarrow (\nu)~^{(2S+1)}n;\rho}$ and likewise a stepwise excitation/ionisation pathway may progress
from lowest metastable $\rho$ via intermediate state $\nu,nS$ to final ionised ion metastable $\nu$ with the effective
$\scriptnew{GCR}$ ionisation coefficient $\scriptnew{SCD}_{\rho \rightarrow (\nu)~^{(2S+1)}n;\nu}$. The population calculation for a given
pathway involves the excited state population structure connecting  the recombining parent $\nu$ and metastable $\rho$. However, an
alternative parent, which is not the intermediate state parent, can be populated by autoionisation of excited states above the
auto-ionisation threshold. That is, if the parent $\nu$ is a metastable and so there exists a lower lying ground state (or possibly
other metastable) of the parent, say $\nu'$. The excited state populations of the $\nu,nS$ system must include such
auto-ionisation processes.   Above the auto-ionisation  threshold, and at low electron densities, auto-ionisation is the dominant
loss mechanism.  However, the bundle-$n$ auto-ionisation transition probabilities scale as $n^{-5}$ and are independent of electron density,
whereas direct ionisation loss rates scale as $n^4$ and vary directly with electron density. For high n-shells, direct ionisation is
the dominant loss process. As the electron  density increases, direct ionisation becomes the dominant loss process for all n-shells
and  auto-ionisation is quenched.  The inclusion of auto-ionisation transition probabilities in  the statistical balance equations
leads to dramatic changes in the population structure as shown in figures \ref{fig:fig3}a, b. In the example, the populations of
$C^{+1} (2s^2~^1S)~^2n$ are built upon a ground state parent so no auto-ionisation pathways are accessible from the excited states.
If the populations are  expressed in terms of the Saha-Boltzmann b-factors, then they show strong overpopulation of  the high
n-shells at low electron density due to dielectronic recombination. As the electron  density is increased, the dielectronic
recombination influence becomes less due to ionisation  of the high n-shell populations with all the b-factors tending  to 1. This
behaviour is typical  for a recombined system built on a ground state parent.  By contrast, the populations of 
$C^{+1}(2s2p~^3P)~^2n$, built upon an excited metastable parent, show powerful depopulation above the auto-ionisation threshold.  As
the electron density increases, the direct ionising  collisions from the excited states compete more strongly and the b-factors tend
to 1. 
\begin{figure}[htp]
\begin{center}
   \begin{minipage}[t]{8.4cm}
   \raggedright
\ \psfig{file=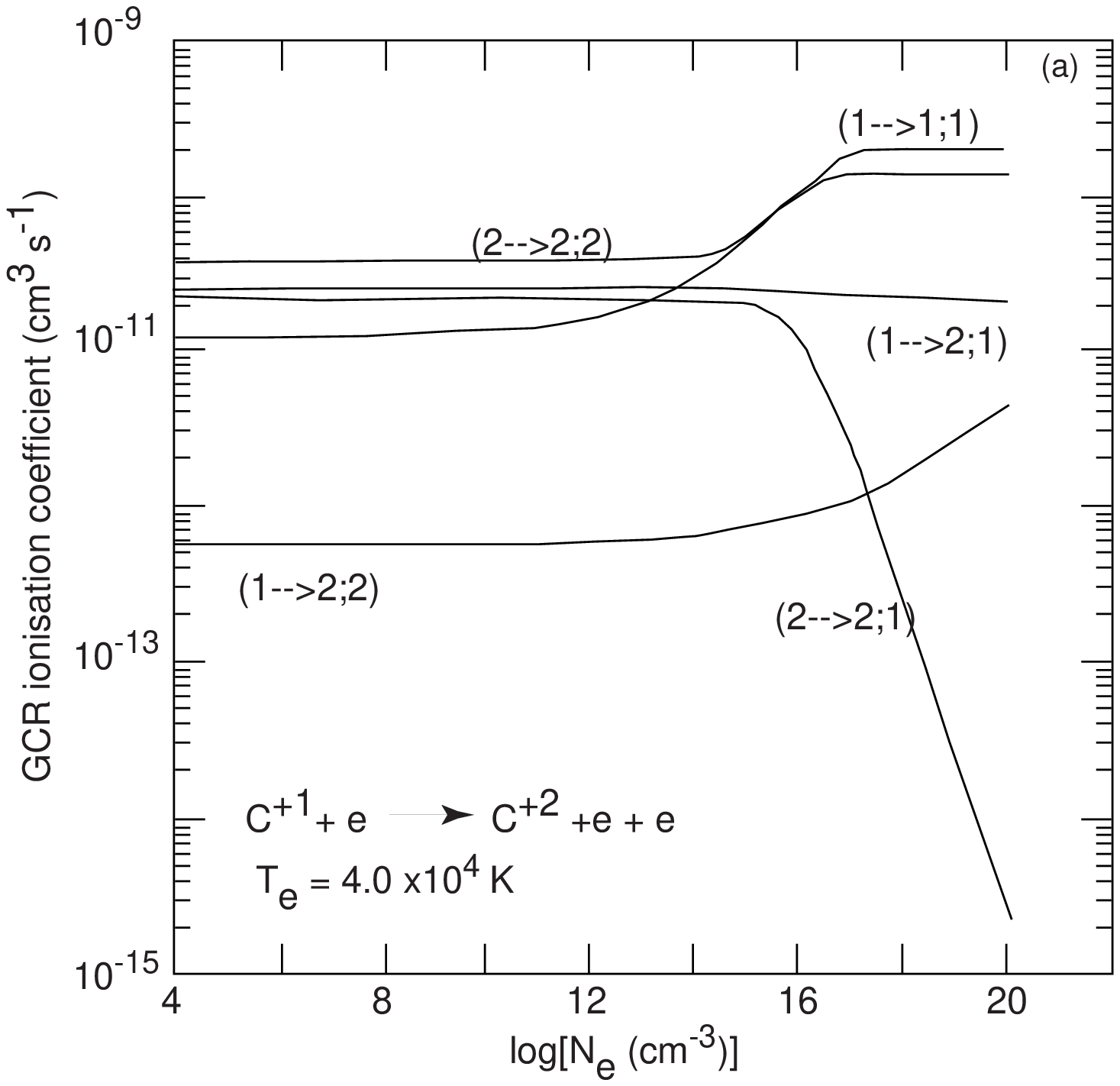,width=8.4cm}
\ \psfig{file=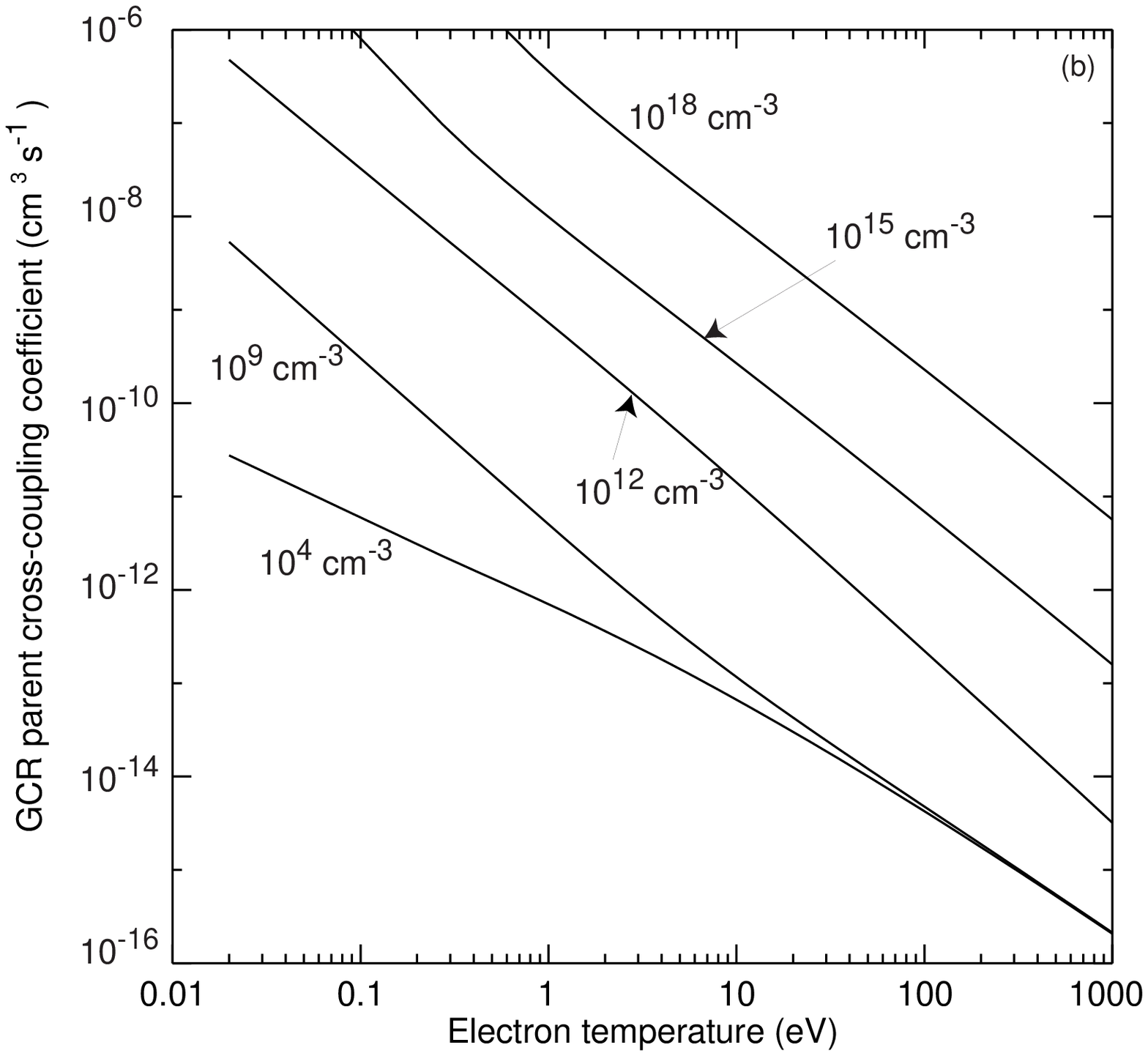,width=8.4cm}
   \end{minipage}
\caption{\label{fig:fig4} Bundle-$nS$ generalised collisional-radiative coefficients. $T_e$ is the electron temperature and $N_e$ the electron
density.  (a) $C^{+1}_{\rho} +e  \rightarrow C^{+2}_{\nu}  + e + e$ for different initial metastable $\rho$, final states $\nu$ and
intermediate parent/spin systems.  Curves are labelled as $\scriptnew{SCD}(\rho \rightarrow (\nu')~^{2S+1}n; \nu)$  where $\rho=1 \equiv
(2s^22p~^2P)$, $\rho=2 \equiv (2s2p^2~^4P)$, $\nu,\nu'=1 \equiv (2s^2~^1S)$ and $\nu,\nu'=2 \equiv (2s2p~^3P)$. (b) $O^{+3}[2s2p^2~^4P] +e
\rightarrow O^{+3}[2s^22p~^2P] + e$ parent cross-coupling coefficients.  Both $O^{+2}[(2s2p^2~^4P)~^3n]$ and $O^{+2}[(2s2p^2~^4P)~^5n]$
recombined systems are included.  The autoionisation  pathway for the former is allowed in LS-coupling, but for the latter proceeds only in
intermediate-coupling (see section \ref{sec:sec2.3}).  Such coefficients follow broadly the behaviour of recombination coefficients.}
\end{center}
\end{figure}

The definitions of generalised collisional-radiative ionisation and recombination coefficients are still relevant but give the
recombination rate from $\nu$ to $\rho$ and the total loss rate from $\rho$ (i.e. with no resolution of final  parent after
ionisation) respectively. The recombination coefficient is already correctly parent/metastable resolved and needs no further
adjustment. The correct parent resolved (partial) ionisation coefficients $\scriptnew{SCD}_{\rho \rightarrow (\nu)~^{(2S+1)}n;\nu'}$  are
derived by constructing the loss vector from each level.  Because the model considers the excited state populations $N_{\nu,nS}$,
direct ionisation only  populates $A^{+z_1}_{\nu}$. The alternative parents are populated by auto-ionisation. The parent resolved
loss vectors are thus given by
\begin{eqnarray}
\label{eqn:eqn36}
	L_{\nu,nS \rightarrow \nu} & = & N_eq_{\nu,nS \rightarrow \nu}	\nonumber \\
	L_{\nu,nS \rightarrow \nu'} & = & A^a_{\nu,nS \rightarrow \nu'}.	
\end{eqnarray}
Ionisation pathways and the expected (partial) $\scriptnew{GCR}$ ionisation coefficients for $C^{+1} +e \rightarrow C^{+2} + e + e$ are summarised in
table \ref{table:tab2}.  From the point of view of recombination from $\nu$ via intermediate states $\nu,nS$ towards $\rho$, autoionisation
allows exit into the alternative parent $\nu'$ before $\rho$ is reached.  This gives the new {\it parent cross coupling coefficient} 
$\scriptnew{XCD}_{\nu \rightarrow (\nu)~^{(2S+1)}n,\nu'}$.  Note that in applications, it is the coefficients summed over intermediate states which
are required as
\begin{equation}
\label{eqn:eqn37}
\scriptnew{XCD}_{\nu \rightarrow \nu'}=\sum_{\nu ,S} \scriptnew{XCD}_{\nu \rightarrow (\nu)~^{(2s+1)}n,\nu'}.
\end{equation}
Behaviours are illustrated in figures \ref{fig:fig4}a, b.   
\begin{table}
\begin{center}
\begin{tabular}{clclc}
Final parent    & Intermediate    	& Initial         & $\scriptnew{GCR}$ ionis.                          & Type  \\
metastable      & system                & metast.         & coefft.                         &       \\
\hline
  1 &		$(2s^2~^1S)~^2n$	&	1	& $\scriptnew{SCD}_{1 \rightarrow (1)~^2n,1}$	& d	\\
  1 &		$(2s2p~^3P)~^2n$	&	1	& $\scriptnew{SCD}_{1 \rightarrow (2)~^2n,1}$	& e-a 	\\
  2 & 					&	1	& $\scriptnew{SCD}_{1 \rightarrow (2)~^2n,2}$	& d	\\
  1 &		$(2s2p~^3P)~^4n$ 	&	2	& $\scriptnew{SCD}_{2 \rightarrow (2)~^4n,1}$	& e-a (IC)\\
  2 &					&	2	& $\scriptnew{SCD}_{2 \rightarrow (2)~^4n,2}$	& d	\\
         
\end{tabular}
\caption{\label{table:tab2} Excited state structures and partial $\scriptnew{GCR}$ ionisation coefficients calculated for the Be-B series. Ionisation 
coefficient nomenclature is $\scriptnew{SCD}_{\rho \rightarrow (\nu)~^{(2S+1)}n,\nu'}$  where $\rho$  indexes the lowest metastable of the
parent/spin system,  with $\nu$ the parent of the spin system and $\nu'$ indexes the final metastable state.  `d' denotes direct ionisation
and `e-a' denotes inner shell excitation-autoionisation contributions.}
\end{center}
\end{table}

\subsection{The low-level + projection model}
\label{sec:sec2.2}

Consider now the set of low levels of an ion.  For $\scriptnew{GCR}$ studies, we are concerned with complete sets of low levels
associated with the valence electron in the ground and excited n-shells.  The span of the low levels is $\Delta
n_{01}=n_1-n_0$ with $n_0$ denoting the ground complex valence n-shell and $n_1$ the highest resolved n-shell.  For
light elements and spectroscopy which extends up to the visible range, we seek $\Delta n_{01}=1$ at minimum and 2
preferably.

The solution for the populations and effective coefficients follow the theory of section \ref{sec:sec1.2.1} and is numerically
straightforward, carried out in the population  representation (called the $p-$representation), rather than $b-$factor or $c-$factor representations (see equation
\ref{eqn:eqn34} and the following lines). The use of the $c-$factor representation, where $c=b-1$, is essential for cancellation error avoidance in computation of
the very high level population structure, but is not necessary for the low levels.  The construction of
the $C_{ij}$ and $r_{i,\nu}$ coefficients is by spline interpolation in electron temperature of data extracted from archives.
Expansion of indirect projection data and its amalgamation with the resolved direct data is by weight matrices as:
\begin{eqnarray}
\label{eqn:eqn38}
C_{i,j}    & = & C^{dir}_{ij}+\sum_{\nu,S}\omega^c_{ij:\nu,S}C^{indir}_{n,n':\nu,S} \nonumber \\
r_{i,\nu}  & = & r^{dir}_i+\sum_S  \omega^r_{i:\nu,S} r^{indir}_{n:\nu,S} \\
L_{i} & = & L^{dir}_{i}+\sum_{\nu,S}\omega^L_{i:\nu,S}L^{indir}_{n:\nu,S} \nonumber 
\end{eqnarray}
where $i \in n$ and $j \in n'$.  Introducing the term statistical weight fractions as 
\begin{equation}
\label{eqn:eqn39}
\omega_{i:n:\nu,S}=\frac{(2S_i+1)(2L_i+1)}{\sum_{j \in n}(2S_j+1)(2L_j+1)}
\end{equation} 
then
\begin{eqnarray}
\label{eqn:eqn40}
\omega^c_{ij:\nu,S} & = &  \omega_{S_{\nu},S}\omega_{i:n:\nu,S}~~~~{\rm for}~n\ne n'  \nonumber \\
 \nonumber \\
\omega^c_{ij:\nu,S} & = &  \omega_{S_{\nu},S}  \nonumber \\
\omega^r_{i:\nu,S} & = & \omega_{i:n:\nu,S} ~~~~~~~~~~~~~{\rm for}~n = n'                      \\
\omega^L_{i:\nu,S} & = & \omega_{S_{\nu},S}  \nonumber 
\end{eqnarray}
and $\omega_{S_{\nu},S}$ is as defined in equation \ref{eqn:eqn33}. Table \ref{table:tab3} illustrates the $C_{ij}$ weighting.

\begin{figure}[htp] 
\begin{center}
   \begin{minipage}[t]{8.4cm}
   \raggedright
\ \psfig{file=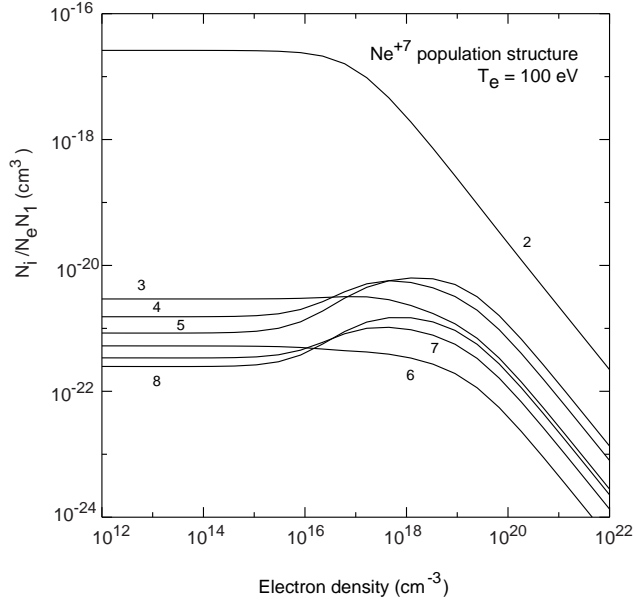,width=8.4cm}
   \end{minipage}
\caption{\label{fig:fig5} Low level population structure of Ne$^{+7}$.  The excitation part driven by the $1s^22s~^2S$ ground state is
denoted by $N_1$. Terms are labelled as 2: $2p~^2P$, 3: $3s~^2S$, 4: $3p~^2P$, 5: $3d~^2D$, 6: $4s~^2S$, 7: $4p~^2P$, 8: $4d~^2D$. At
low density, the population ratios $N_i/N_eN_1$ become flat tending to the coronal values.  At high density the population ratios
decrease inversely with the electron density, $N_e$, tending to the Saha-Boltzmann values.  The intermediate region is the
collisional-radiative regime.}
\end{center}
\end{figure}

In the deduction of spectral emission coefficients between low levels there can be some confusion.  Following the definition of
equation \ref{eqn:eqn21}, in the resolved low level picture, the emission coefficient is referred to a particular metastable.  If
metastables are neglected, so that there is only a ground state and all other levels are viewed as excited, the reference is to
the ground state.  On the other hand, a metastable treated as an ordinary excited level will have a quasistatic population
comparable to that of the ground so that $\sum_i N_i = N_{tot} \ne N_1$ and $\scriptnew{PEC}_{tot,j \rightarrow
k}=\scriptnew{PEC}_{1,j \rightarrow k}N_1/N_{tot}$ and this is the coefficient which should be used with a `stage-to-stage' (that
is un-generalised picture) ionisation balance.      
\begin{table}[h]
\begin{center}
\begin{tabular}{clcccll}
Index    & Term    	    & Spin         & Parent     & Shell      &\multicolumn{2}{c}{n-shell weights}  \\
         &        	    &              &            &            & n=2    & n=3     \\
\hline
  1      &  $ 2s^22p~^2P$   &      2        &  1         &	2    & 0.25    &  -	\\
  1      &  $ 2s^22p~^2P$   &      2        &  2         &	2    & 0.25    &  -	\\
  2      &  $ 2s2p^2~^4P$   &      4        &  2         &	2    & 0.786   &  -	\\
  3      &  $ 2s2p^2~^2D$   &      2        &  1         &	2    & 0.45    &  -	\\
  3      &  $ 2s2p^2~^2D$   &      2        &  2         &	2    & 0.45    &  -	\\
  4      &  $ 2s2p^2~^2S$   &      2        &  1         &	2    & 0.05    &  -	\\
  4      &  $ 2s2p^2~^2S$   &      2        &  2         &	2    & 0.05    &  -	\\
  5      &  $ 2s2p^2~^2P$   &      2        &  2         &	2    & 0.25    &  -	\\
  5      &  $ 2s2p^2~^2P$   &      2        &  2         &	2    & 0.25    &  -	\\
  6      &  $ 2s^23s~^2S$   &      2        &  1         &	3    &  -      & 0.0667	\\
  6      &  $ 2s^23s~^2S$   &      2        &  2         &	3    &  -      & 0.0667 \\
  7      &  $ 2s^23p~^2P$   &      2        &  1         &	3    &  -      & 0.3333 \\
  7      &  $ 2s^23p~^2P$   &      2        &  2         &	3    &  -      & 0.3333 \\
  8      &  $ 2p^3~^4P$     &      4        &  2         &	3    & 0.214   &  -	\\
  9      &  $ 2s^23d~^2D$   &      2        &  1         &	3    &  -      & 0.6    \\
  9      &  $ 2s^23d~^2D$   &      2        &  2         &	3    &  -      & 0.6    \\
  10      &  $ 2s2p3s~^4P$   &      4        &  2         &	3    &  -      & 0.1089 \\
  11      &  $ 2s2p3p~^4D$   &      4        &  2         &	3    &  -      & 0.1881 \\
  12      &  $ 2s2p3p~^4S$   &      4	     &  2	  &	3    &  -      & 0.0297 \\
  13      &  $ 2s2p3p~^4P$   &      4	     &  2	  &	3    &  -      & 0.1089 \\
  14      &  $ 2s2p3d~^4F$   &      4	     &  2	  &	3    &  -      & 0.2673 \\
  15      &  $ 2s2p3d~^4D$   &      4	     &  2	  &	3    &  -      & 0.1881 \\
  16      &  $ 2s2p3d~^4P$   &      4        &  2         &	3    &  -      & 0.1089 \\
\end{tabular}
\caption{\label{table:tab3} Fractionising of bundle-$nS$ matrix projection onto low levels for B-like ions with two resolved
n-shells.}
\end{center}
\end{table}
Figure \ref{fig:fig5} illustrates the behaviour of low level populations.  The graph is of the parameter $\scriptnew{F}^{(exc)}_{i1}$
from equation \ref{eqn:eqn20}. 

\subsection{Specific reactions}
\label{sec:sec2.3}
Energy levels for high bundle-$nS$ levels and their A-values, Maxwell averaged collision strengths, radiative recombination coefficients,
dielectronic recombination  coefficients and ionisation coefficients are generated from a range of parametric formulae and
approximations described in earlier works (Burgess and Summers, 1976; Summers, 1977; Summers and Hooper, 1983; Burgess and Summers,
1987; Summers, 2004).  High quality specific data when available are acccessed from archives and substituted for the default
values.  This is a systematic procedure for dielectronic coefficients (see sections \ref{sec:sec2.3.1} and \ref{sec:sec5} below).  

For low levels, a complete basis of intermediate coupling energy level, A-value and scaled Born approximation collision data, spanning the principal quantum shell
range $n_0 \leq n \leq n_1$ is generated automatically using the Cowan (1981) or Autostructure (Badnell, 1986)  procedures.  This is called our {\it baseline}
calculation.  These data are merged with more restricted (in level coverage) but similarly organised highest quality data from archives where available (e.g.
R-matrix data such as Ramsbottom \etal (1995)).   The data collection is compressed by appropriate summing and averaging to form a complete LS term basis and
augmented with comprehensive high quality LS resolved dielectronic recombination, radiative recombination and collisional ionisation coefficients mapped from
archives (see sections \ref{sec:sec2.3.1}, \ref{sec:sec2.3.2} and \ref{sec:sec5} below).  The radiative data has its origin in the work of Burgess and Summers
(1987).  The final data collection is called a {\it specific ion file} (ADAS data format {\it adf04}).  The detailed content is examined in section
\ref{sec:sec3}.   Details of state selective dielectronic recombination and ionisation coefficients are given in the following sub-sections.             

\subsubsection{State-selective dielectronic recombination}
\label{sec:sec2.3.1}

State selective dielectronic recombination coefficients are required to all resolved low levels and to all bundle-$nS$ shells for
the various initial and intermediate state metastable parents $\nu$ for $\scriptnew{GCR}$ modelling.  These are very extensive
data and have been prepared for the present $\scriptnew{GCR}$ work through an associated international `DR Project' summarised in
Badnell \etal (2003) [hereafter called {\it DR - paper I}] and detailed in subsequent papers of the series .  Based on the
independent particle, isolated resonance, distorted wave (IPIRDW) approximation, the partial dielectronic recombination rate
coefficient $\alpha^{(d)}_{\nu \rightarrow i}$ from an initial metastable state $\nu$ into a  resolved final state $i$ of an ion
$\scriptnew{A}^{+z}$ is given by
\begin{eqnarray}
\label{eqn:eqn41}
\alpha^{(d)}_{\nu \rightarrow i}&=&\left({4\pi a^2_0 I_\rmH \over k T_\rme}\right)^{3/2} \sum_{p,j}{\omega_{p,j} 
\over 2\omega_{\nu}}\,\rme^{-E_c/k T_\rme}\nonumber \\
& & \times{ \sum_{l}A^{\rma}_{p,j \rightarrow \nu, E_cl} \, A^{\rmr}_{p,j \rightarrow i} 
\over \sum_{h} A^{\rmr}_{p,j \rightarrow h} + \sum_{m,l} A^{\rma}_{p,j \rightarrow m, E_cl}}\, 
\end{eqnarray}
where $\omega_{p,j}$ is the statistical weight of the $(N+1)$-electron doubly-excited resonance state $j$, $\omega_\nu$ is the statistical weight of the
$N$-electron target state and the autoionization ($A^\rma$) and radiative ($A^\rmr$) rates are in inverse seconds. The suffix $p$ is used here to denote a parent
ion state. $E_c$ is the energy of the continuum electron (with angular momentum $l$), which is fixed by the position of the resonances. Note that the parent states
$p$ are excited, that is they exclude the metastable parents $\nu'$.  The code {\sc autostructure} (Badnell, 1986; Badnell and Pindzola, 1989; Badnell, 1997) was
used to calculate multi-configuration LS and intermediate coupling energy  levels and rates within the IPIRDW approximation. The code makes use both of
non-relativistic and semi-relativistic wavefunctions (Pindzola and Badnell, 1990) and is efficient and accurate for both the resolved low level and high-$n$ shell
problems.  Lookup tables (see section \ref{sec:sec5}; {\it adf09}) are prepared comprising state selective recombination coefficients at a standard set of
$z$-scaled temperatures, for each metastable parent, to all LS resolved terms of the recombined ion with valence electron up to $n$-shell $n_1^{(d)} \ge n_1$
(normally $n_1^{(d)}=5$) and to bundle-$nS$ levels of a representative set of $n$-shells (usually spanning $n_0$ to 999).  These bundle-$nS$ coefficients are
simple sums over orbital states $l \in n$ and so apply at zero density. This provides an extensive, but still economical, tabulation. 

\begin{figure}[htp] 
\begin{center}
   \begin{minipage}[t]{8.4cm}
   \raggedright
\ \psfig{file=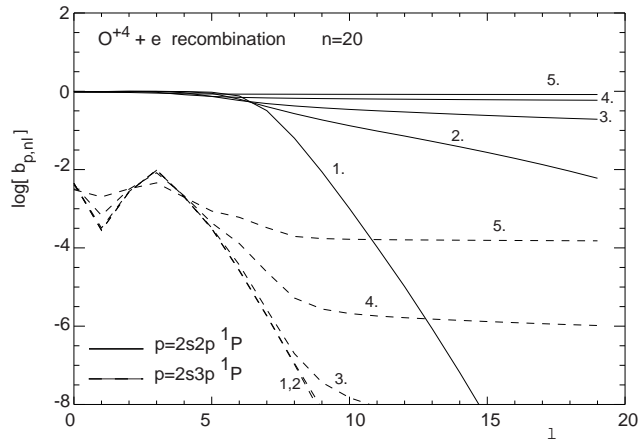,width=8.4cm}
   \end{minipage} \caption{\label{fig:fig6} O$^{4+}$ dielectronic recombination. $b_{p,nl}$ factors as a function of
outer electron orbital angular momentum $l$ for doubly-excited states of O$^{3+}$ relative to O$^{4+}$
$2\rms^2~^1\rmS$ for $p= 2\rms2\rmp~^1\rmP$ and $2\rms3\rmp~^1\rmP$, $n=20$, $T_e=\rm{10}^6$ K, $N_e=N_p$,
$Z_{eff}=1$. Cases: 1. $N_\rme=\rm{10}^{10}$ cm$^{-3}$; 2. $N_\rme=\rm{10}^{12}$ cm$^{-3}$; 3. $N_\rme=\rm{10}^{13}$
cm$^{-3}$; 4. $N_\rme=\rm{10}^{14}$ cm$^{-3}$; 5. $N_\rme=\rm{10}^{15}$ cm$^{-3}$. Note the alternative Auger channel
reduction for the $p= 2\rms3\rmp~^1\rmP$ graphs.}
\end{center}
\end{figure}
In {\it DR paper-I}, we introduced  an associated code, the `Burgess--Bethe general program' BBGP.  BBGP is used here as a
support function in a model for the $l$-redistribution of doubly-excited states which provide a correction to the accurate, but
unredistributed, dielectronic data.  The redistributed data (regenerated in the same {\it adf09} format) are normalised to IPIRDW
at zero density.  The procedure is similar to that for singly excited systems. For $LS$-averaged levels, the number
densities expressed in terms of their deviations, $b_{p,nl}$ from Saha-Boltzmann, and referred to the initial parent $\nu'$, are
given by
\begin{eqnarray}
\label{eqn:eqn42}
N_{p,nl} = N_\rme N^{+}_{\nu'}8\left[\frac{\pi a_0^2I_\rmH}{kT_\rme}\right]^{3/2} 
\frac{\omega_{p,nl}}{\omega_{\nu'}}e^{-E/kT_\rme}b_{p,nl}\, .
\label{eq22}
\end{eqnarray}
\noindent Thus, in the BBGP  zero-density limit, with only resonant capture from the
$\nu$ parent balanced by Auger breakup and radiative stabilization back to the same parent, we have
\begin{eqnarray}
\label{eqn:eqn43}
b_{p,nl} & = & \left(\frac{\sum_{l'}A^\rma_{p,nl \rightarrow \nu'k'l'}}{\sum_{l'}A^\rma_{p,nl 
\rightarrow \nu k'l'}+A^\rmr_{p,nl \rightarrow \nu',nl}}\right) \, .   
\end{eqnarray}
In the extended BBGP model, we include resonant-capture from initial metastables other
than the ground, dipole-allowed collisional redistribution between adjacent doubly-excited
$l$-substates of the same $n$ by secondary ion- and electron-impact, and losses by `alternate' Auger break-up
and parent radiative transition pathways.  The population equations for the $l$-substates of
a doubly-excited $n$-shell become   
\begin{eqnarray}
\label{eqn:eqn44}
&&-\left(N_{\rm e} q^{(\rm e )}_{nl-1 \rightarrow nl}+N_{z_{eff}}q^{(z_{eff})}_{nl-1 \rightarrow nl}\right)
N_{p,nl-1}\nonumber \\
&&+\left(\sum_{l'=l\pm1}N_{\rm e} q^{(\rm e )}_{nl \rightarrow nl'}
+\sum_{l'=l\pm1}N_{z_{eff}}q^{(z_{eff})}_{nl \rightarrow nl'} \right.\nonumber \\
&&+\left.\sum_{p_1=1}^{p-1}\sum_{l'=l-1}^{l+1}A^{\rm a}_{p,nl \rightarrow p_1,\kappa
 l'}+\sum_{p_1=1}^{p-1}A^{\rm r}_{p,nl \rightarrow p_1,nl}\right)N_{p,nl}\nonumber \\
&& -\left(N_{\rm e} q^{(\rm e )}_{nl+1 \rightarrow nl}+N_{z_{eff}}q^{(z_{eff})}_{nl+1 \rightarrow nl}\right)
N_{p,nl+1} \nonumber \\
&&= N_{\rm e}\sum_{\nu=1}^{M}\sum_{l'=l-1}^{l+1}q^{\rm c}_{\nu,\kappa l' \rightarrow p,nl}N_{\nu} 
+\sum_{p_1=p+1}^{P}A^{\rm r}_{p_1,nl \rightarrow p,nl}N_{p_1,nl}\, .
\end{eqnarray}
$q^c$ denotes resonance capture coefficients. $M$ denotes the number of parent metastables which are the starting point for resonance capture whereas $P$ denotes
the number of true excited parent states. Ion impact redistributive collisions are effective and are represented in
the equations as $z_{eff}$ ion contributions. The density corrected bundle-$nS$ recombination coefficients are then
\begin{equation}
\label{eqn:eqn45}
\alpha^{(d,IPIRDW)}_{\nu' \rightarrow \nu',nS}(N_e) =\alpha^{(d,IPIRDW)}_{\nu' \rightarrow \nu',nS}(N_e=0) \frac{\sum_l
\alpha^{(d,BBGP)}_{\nu' \rightarrow \nu',nl}(N_e)}{\sum_l
\alpha^{(d,BBGP)}_{\nu' \rightarrow \nu',nl}(N_e=0)}.  
\end{equation}
Details are given in {\it DR paper-I}.  Figure \ref{fig:fig6} shows the effects of redistribution of the doubly excited
$b_{p,nl}$ factors in the recombination of $O^{+4}$.

Autoionisation rate calculations in LS coupling exclude breakup of non-ground metastable parent based `singly excited' systems by
spin change.  Such breakup only occurs in intermediate coupling but cannot be ignored for a correct population assessment even in
light element ions.  These rates are computed in separate intermediate coupling dielectronic calculations.  These rates
are main contributors to the parent metastable cross-coupling identified earlier (cf. the $O^{+2}[(2s2p^2~^4P)~^5n]$
intermediate states in figure \ref{fig:fig4}b leading back to the $O^{+3}(2s^22p~^2P)$ parent).

\subsubsection{State-selective ionisation}
\label{sec:sec2.3.2}

Direct ionisation coefficients for excited n-shells in bundle-$nS$ population modelling are evaluated in the
exchange-classical-impact parameter (ECIP) approximation (Burgess and Summers, 1976).  The method which merges a
symmetrised classical binary encounter model with an impact parameter model for distant encounters is relatively
simple and of moderate precision, but has a demonstrated consistency and reliability for excited n-shells in
comparison with more eleborate methods.  There is a higher precision requirement for state selective ionisation from 
metastable and low levels.  Such ionisation includes direct and excitation/autoionisation parts, but the latter only
through true excited parents. Stepwise ionisation is handled in the collisional-radiative population models.  There
are extensive ionisation cross-section measurements, but these are in general unresolved and so are of value
principally for renormalising of resolved theoretical methods.  The most powerful theoretical methods used for
excitation, namely R-matrix with pseudostates (RMPS) (Bartschat and Bray, 1996; Ballance \etal, 2003) and
convergent-close-coupling (CCC) (Bray and Stelbovics, 1993), have some capability for ionisation, but at this stage
are limited to very few electron systems as is the time-dependent-close-coupling (TDCC) (Pindzola and Robicheaux,
1996; Colgan \etal, 2003) ionisation method.  For $\scriptnew{GCR}$ calculations, we have relied on the procedures of
Summers and Hooper (1983) for initial and final state resolution of total ionisation rate coefficients and on the
distorted wave approximation.  The distorted wave method is our main method for extended $\scriptnew{GCR}$ studies for
many elements and most stages of ionisation. RMPS and TDCC studies are directed mostly at the neutral and near neutral
systems where the distorted wave method is least reliable (Loch \etal, 2005).  We use the configuration average
distorted ({\it CADW}) wave approach of Pindzola, Griffin and Bottcher (1986).  It has reasonable economy of
computation while allowing access to complex, multi-electron ions, highly excited states, excitation/autoionisation
and radiation damping.  It expresses the configuration averaged excitation cross-section for 
\begin{equation}
\label{eqn:eqn46}
(n_1l_1)^{q_1+1}(n_2l_2)^{q_2-1}\bar{k}_il_i \rightarrow (n_1l_1)^{q_1}(n_2l_2)^{q_2}\bar{k}_fl_f \nonumber
\end{equation}
as                 
\begin{eqnarray}
\label{eqn:eqn47}
\bar{\sigma}_{excit} & = & \frac{8\pi}{\bar{k}_i^3\bar{k}_f}(q_1+1)(4l_2+3-q_2) \nonumber \\
& & \sum_{l_i,l_f}(2l_i+1)(2l_f+1)M_{2f;1i}
\end{eqnarray}
and the configuration averaged ionisation cross-section for
\begin{equation}
\label{eqn:eqn48}
(n_1l_1)^{q_1+1}\bar{k}_il_i \rightarrow (n_1l_1)^{q_1}\bar{k}_el_e\bar{k}_fl_f \nonumber
\end{equation}
as                 
\begin{eqnarray}
\label{eqn:eqn49}
\bar{\sigma}_{ionis} & = & \frac{32\pi}{\bar{k}_i^3\bar{k}_e\bar{k}_f}(q_1+1) \nonumber \\
& & \sum_{l_i,l_e,l_f}(2l_i+1)(2l_e+1)(2l_f+1)M_{ef;1i}
\end{eqnarray}
where $M_{2f;1i}$ are squared two-body Coulomb matrix elements,  $\bar{\sigma}$ denotes the average cross-section and $\bar{k}_i$, $\bar{k}_f$ and
$\bar{k}_e$ denote average initial, final and ejected electron momenta respectively in the configuration average picture.  The configuration
average direct ionisation cross-sections and rate coefficients may be unbundled back to resolved form using angular factors obtained by Sampson
and Zhang (1988). Note that the excitations described here are to auto-ionising levels and so resolved Auger yields may be used which are the same
as those in the dielectronic calculations of section \ref{sec:sec2.3.1} above.  In fact the ionisation coefficient calculation results are
structured and archived in ADAS (format {\it adf23}) in a manner very similar to state selective dielectronic recombination. Extensive studies
have been carried out on light and heavy elements (Colgan \etal, 2003; Loch \etal, 2003).

\section{Fundamental atomic data for low levels of ions}
\label{sec:sec3}

For bundle-$nS$ modelling, the expected fundamental data precision is $\lesssim$ 30\%  for excitation and ionising collisional rate
coefficients, $\lesssim$ 10\% for A-values and state selective recombination coefficients and $\lesssim$  1\% for energies.

\begin{table}
\begin{center}
\begin{tabular} {cccccccc}
Ion		& E & A   & \multicolumn{2}{c}{$\Upsilon_{ij}$}	    & $\alpha^{(r)}$ & $\alpha^{(d)}$ & $S$ \\
		& . &	  &  $\Delta n=0$      & $\Delta n>0$         & 	         &	      &     \\
\hline
${\rm C}^0   $  & a & b   &   	b	     &       b,c  	    &	  b	 &     b      &  b \\ 
${\rm C}^{+1}$  & a & b   &   	b	     &       b,c  	    &	  b	 &     b      &  b \\ 
${\rm C}^{+2}$  & a & b   &   	a,b	     &       b,c  	    &	  b	 &     b      &  b \\ 
${\rm C}^{+3}$  & a & a,b &   	b	     &       b,c  	    &	  b	 &     b      &  b \\ 
${\rm C}^{+4}$  & a & a   &   	b	     &       b,c  	    &	  b	 &     b      &  b \\ 
${\rm C}^{+5}$  & a & a   &   	b	     &       b  	    &	  b	 &     b      &  b \\ 
${\rm O}^0   $  & a & b,c &   	b	     &       b  	    &	  b	 &     b      &  b \\ 
${\rm O}^{+1}$  & a & b,c &   	b	     &       b  	    &	  b	 &     b      &  b \\ 
${\rm O}^{+2}$  & a & b,c &   	a,b	     &       b  	    &	  b	 &     b      &  b \\ 
${\rm O}^{+3}$  & a & b   &   	a	     &       a  	    &	  b	 &     b      &  b \\ 
${\rm O}^{+4}$  & a & b   &   	a	     &       a  	    &	  b	 &     b      &  b \\ 
${\rm O}^{+5}$  & a & b   &   	b	     &       b  	    &	  b	 &     b      &  b \\ 
${\rm O}^{+6}$  & a & a   &   	b	     &       b,c  	    &	  b	 &     b      &  b \\ 
${\rm O}^{+7}$  & a & a   &   	b	     &       b,c  	    &	  b	 &     b      &  b \\ 
${\rm Ne}^0  $  & a & b,c &   	b	     &       c  	    &	  b	 &     b      &  b \\ 
${\rm Ne}^{+1}$ & a & c   &   	b	     &       b  	    &	  b	 &     b      &  b \\ 
${\rm Ne}^{+2}$ & a & b   &   	a	     &       b  	    &	  b	 &     b      &  b \\ 
${\rm Ne}^{+3}$ & a & b   &   	a	     &       b  	    &	  b	 &     b      &  b \\ 
${\rm Ne}^{+4}$ & a & b   &   	a	     &       b  	    &	  b	 &     b      &  b \\ 
${\rm Ne}^{+5}$ & a & b   &   	a	     &       b  	    &	  b	 &     b      &  b \\ 
${\rm Ne}^{+6}$ & a & b   &   	a	     &       b  	    &	  b	 &     b      &  b \\ 
${\rm Ne}^{+7}$ & a & b   &   	b	     &       b  	    &	  b	 &     b      &  b \\ 
${\rm Ne}^{+8}$ & a & a   &   	b	     &       c  	    &	  b	 &     b      &  b \\ 
${\rm Ne}^{+9}$ & a & a   &   	b	     &       c  	    &	  b	 &     b      &  b \\ 
\end{tabular}
\caption{\label{table:tab4}Fundamental data precision categorising for the ions of carbon, oxygen and neon. The definition of the
categories and justification of the categories for ${\rm Ne}^{+6}$ are given in the text.}
\end{center}
\end{table}

For low level modelling many sources are used. A rating is given for the classes of fundamental data for the ions of carbon, oxygen
and neon in table \ref{table:tab4} which is based on the following considerations.  For energy levels, categories are {\it a}
spectroscopic,  {\it b} $\lesssim$ 0.5\% and {\it c} $\lesssim$ 1.0\%.  Category {\it c} is anticipated from ab initio
multi-configuration structure calculations, {\it b} from such calculations with extended optimising and {\it a} reflects direct
inclusion of experimental energies from reference sources.  For A-values, categories are {\it a} $\lesssim$ 5\%,  {\it b}
$\lesssim$ 10\% and {\it c} $\lesssim$ 25\%.  Category {\it c} is anticipated from our baseline calculations, {\it b} from optimised
multi-configuration structure calculations with extended optimising and {\it a} from specific studies in the literature.  For
electron impact Maxwell averaged collision strengths, $\Upsilon$, the categories are {\it a} $\lesssim$ 10\%,  {\it b} $\lesssim$
20\% and {\it c} $\lesssim$ 35\%.  Category {\it c} is from our baseline calculations, {\it b} from distorted wave calculations and
{\it a} from specific R-matrix calculations, equivalent methods or experiment. For radiative recombination, the categories are {\it
a} $\lesssim$ 5\%,  {\it b} $\lesssim$ 10\% and {\it c} $\lesssim$ 50\%.  Category {\it c} is from scaled methods using
hydrogenic matrix elements, {\it b} from distorted wave one-electron wave functions in an optimised potential using spectroscopic
energies and {\it a} from specific R-matrix calculations and experiment. Category {\it b} is the baseline in ADAS. For dielectronic
recombination, the categories are {\it a} $\lesssim$ 20\%,  {\it b} $\lesssim$ 30\% and {\it c} $\lesssim$ 45\%.  Category {\it
c} is from BBGP approximations, {\it b} from LS-coupled calculations using Autostructure from the DR Project and {\it a} from
IC-coupled calculations using Autostructure with parent and lowest resonance energy level adjustments from the DR Project. Category
{\it b} is the baseline in ADAS. It is to be noted that the theoretical relative precision which is consistent with the variation between the
three categories is 15\% better but dielectronic recombination comparisons with experiment still show unexplained discrepancies at the
20\% level so the present categories are safe.  For ionisation, the categories are {\it a} $\lesssim$ 10\%,  {\it b} $\lesssim$ 25\% and {\it
c} $\lesssim$ 40\%.  Category {\it c} is from ECIP approximations, {\it b} from configuration average with angular factor term
resolution and {\it a} from RMPS, TDCC calculations and experiment. Category {\it b} is the baseline in ADAS.         

As indicated in section \ref{sec:sec2.3}, these various data are assembled in an {\it adf04} file which is sufficient to support the
primary low-level population calculation.  The tabulations are at a set of temperatures arising from a fixed set of z-scaled
temperatures (see section \ref{sec:sec2.1.1}) which spans the full range to asymptotic regions of reaction data.  Collision data are
converted to these ranges using {\it C-plots} (Burgess and Tully (1992) and this procedure also flags data errors or queries. The
precision of the specific ion file determines the achievable precision of all derived populations, emissivities and
collisional-radiative coefficients (Whiteford \etal, 2005; O'Mullane \etal, 2005). The assembling of data
in the {\it adf04} is systematic and orderly.  A comment section at the end of the file details the assembly steps, implementer,
codes used and dates.  This includes baseline and supplementation files, merging, LS compression, dielectronic recombination data
inclusion etc.  Also there is extended detail of orginal data sources and a history of updates. A given {\it adf04} file represents
a `snapshot' of the state of available knowledge at the time.  It is subject to periodic review and ADAS codes (see section
\ref{sec:sec5}) are designed to enable easy reprocessing of all derived data following fundamental data update.  The grading for each
ion given in table \ref{table:tab4} is justified and supported by the comments from its {\it adf04} file.   A summary from ${\rm
Ne}^{+6}$ is given in the following paragraphs in illustration. 
\begin{figure}[htp] 
\begin{center}
   \begin{minipage}[t]{8.4cm}
   \raggedright
\ \psfig{file=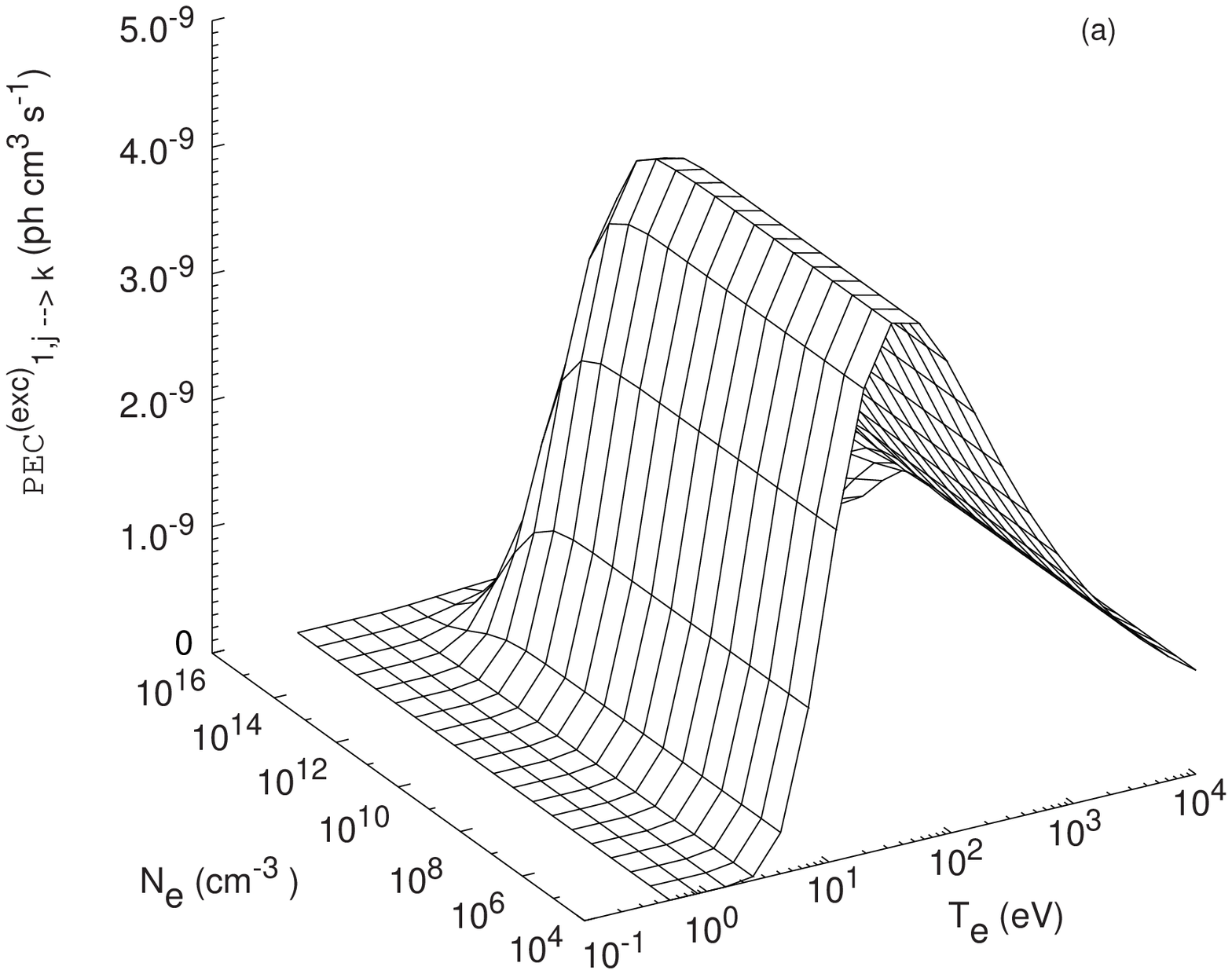,width=8.4cm}
\ \psfig{file=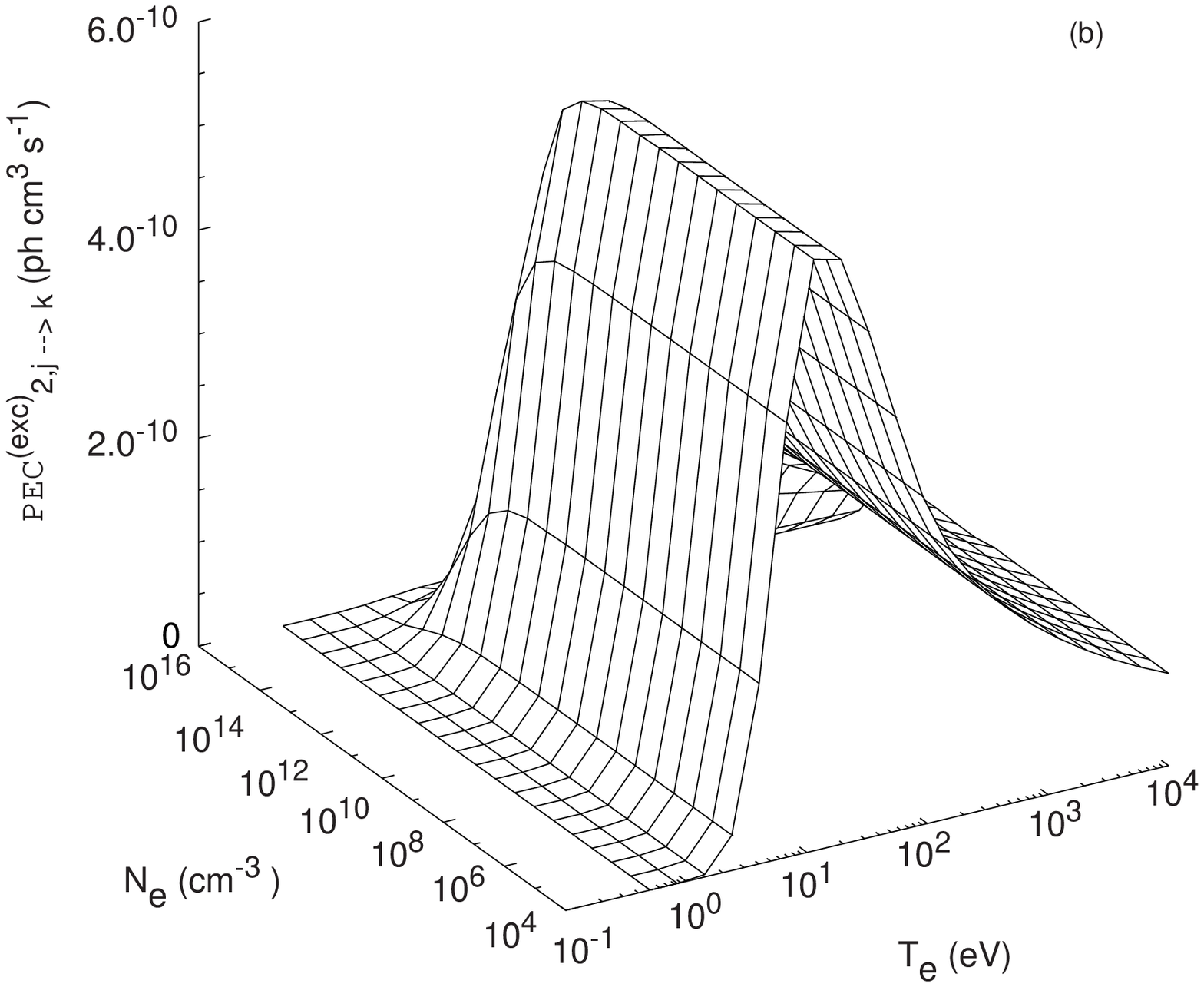,width=8.4cm}
\ \psfig{file=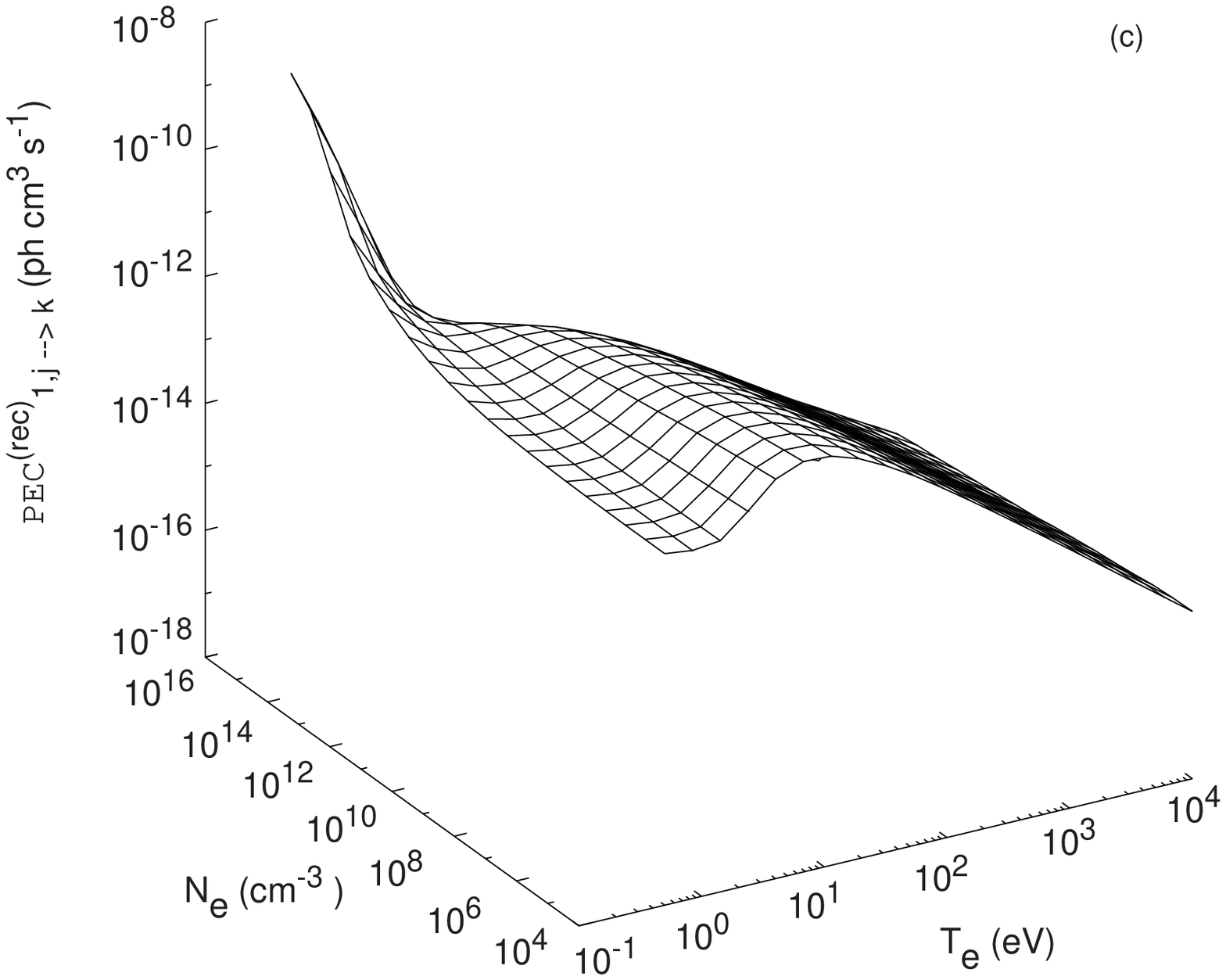,width=8.4cm}
   \end{minipage}
\caption{\label{fig:fig7} Excitation and recombination photon emissivity functions, ${\scriptnew {PEC}}$s, driven by the ground
metastables, are shown for the C II 858~\AA~spectrum line. The coefficients depend on both electron temperature and electron
density. The $(exc)$ part decreases at high electron density due principally to stepwise ionisation losses from the upper level
of the transition.  The behaviour of the $(rec)$ part shows both suppression of dielectronic recombination at moderate density
and then enhancement due to three-body recombination at high density and low temperature.}
\end{center}
\end{figure}
For ${\rm Ne}^{+6}$, the low levels span 44 terms, including up to the $n=5$ shell built on the parent $1s^22s~^1S$ and $n=3$ shell
built on the $1s^22p~^1S$.  Only the $1s^22s~^1S$ parent is treated as a metastable from the $\scriptnew{GCR}$ point of view.  The
intermediate coupled baseline data set is  {\it copmm$\#$10-ic$\#$ne6.dat} with preferred supplementary energy  A-value and $\Upsilon$
data merged from {\it copjl\#be-ic$\#$ne6.dat}.  Ionisation potentials and energy levels are from the NIST Standard Reference Database
apart from $2p3p~^1S$ (Kelly, 1987) and $2p3d~^3F$ levels by (Ramsbottom \etal, 1995).  The categorisation is {\it a}.  A-values were
drawn from the Opacity Project( Opacity  Project Team, 1996; Tully \etal, 1991 ) as justified (for N IV and O V) by Wiese, Fuhr and
Deters (1996) and supported/adjusted  by $\lesssim$ 3\% by Fleming \etal (1996a,b), J\"{o}nsson \etal (1998),  Froese Fischer, Godefeid and
Olsen (1997), Froese Fischer, Gaigalas and Godefried(1997), Nussbaumer and Storey (1979), Sampson, Goett and Clark (1984) and
Ramsbottom \etal (1995).  Category {\it b} is safe.  $\Upsilon$s are taken from Ramsbottom \etal (1995). These are for a 26 LS
eigenstate multichannel R-matrix calculation.  The categories assigned are {\it a} and {\it b}.  Radiative and dielectronic
recombination and ionisation all follow the baselines in ADAS, that is categories {\it b} although the summed and averaged ionisation
rate coefficient (over metastables) is normalised to experiment.

\section{Illustrative results}
\label{sec:sec4}
\begin{figure}[htp]
\begin{center}
   \begin{minipage}[t]{8.4cm}
   \raggedright
\ \psfig{file=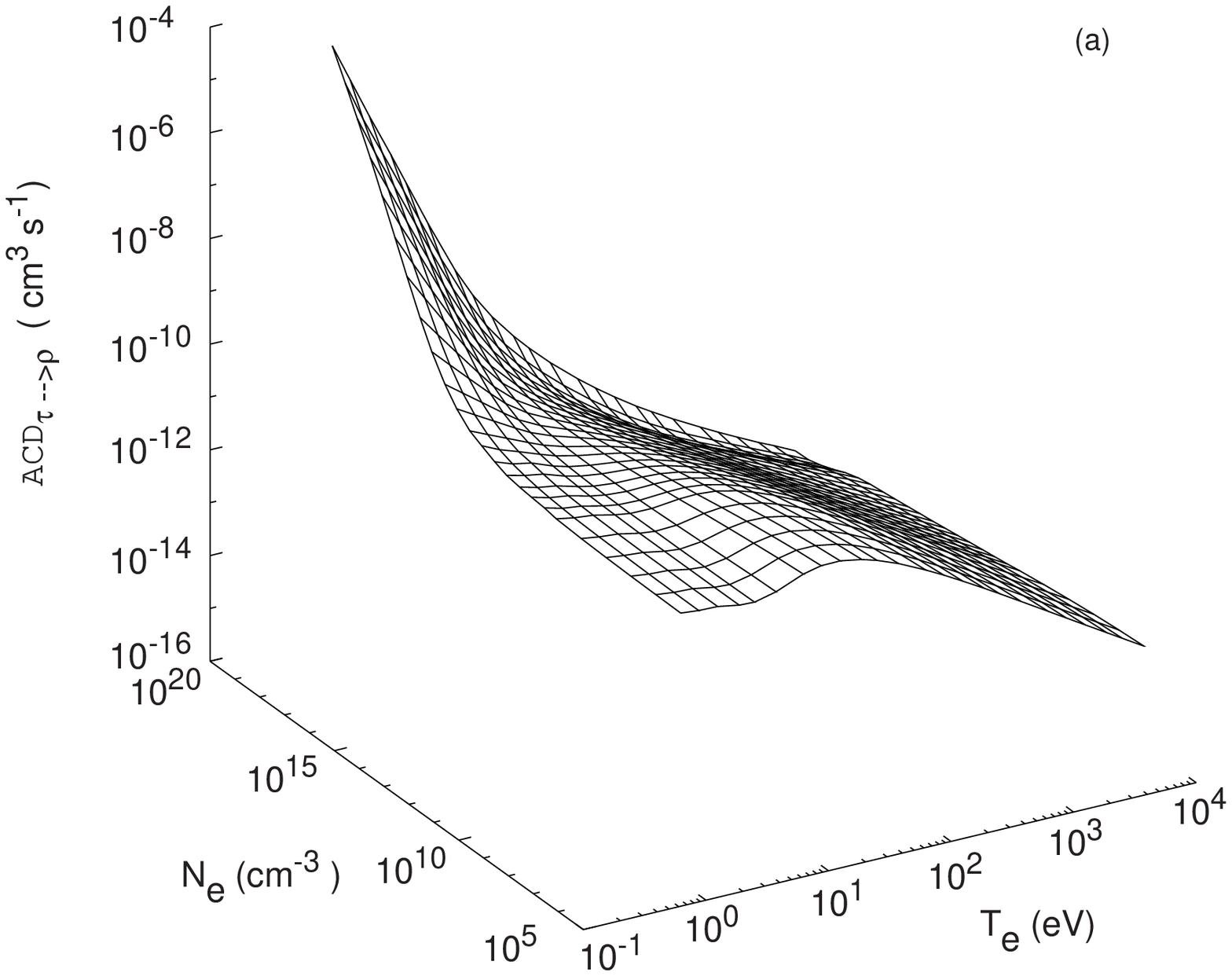,width=8.4cm}
\ \psfig{file=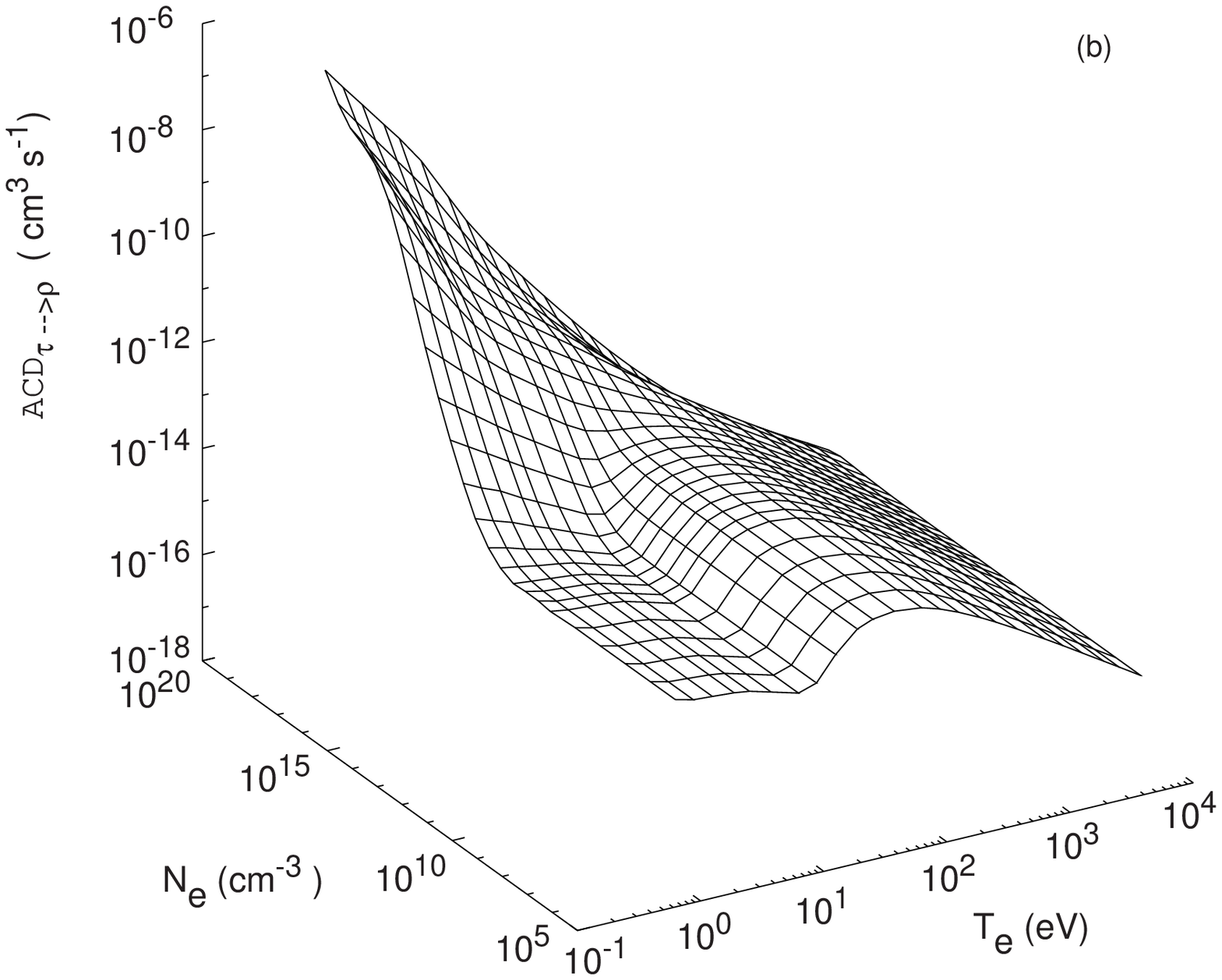,width=8.4cm}
\ \psfig{file=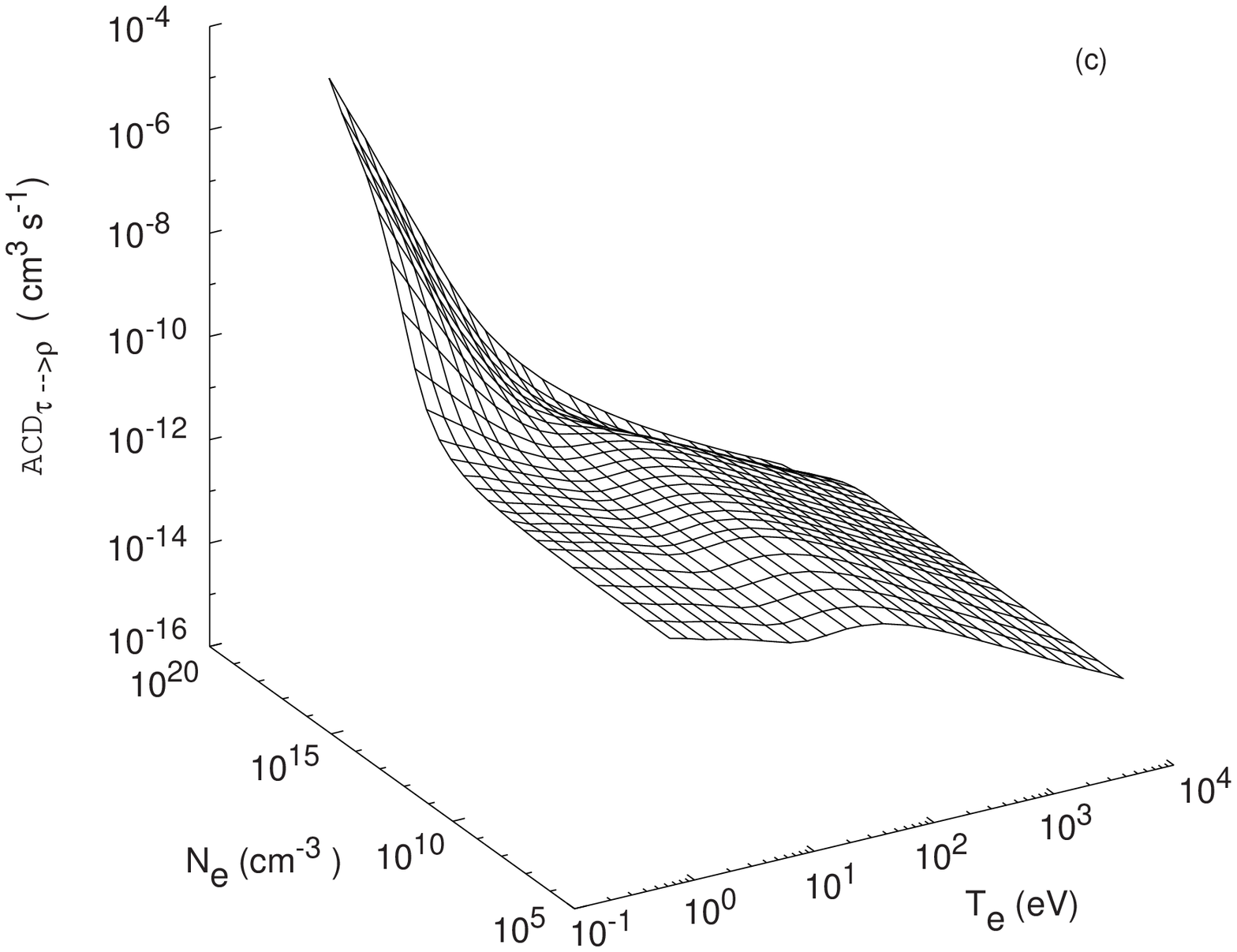,width=8.4cm}
   \end{minipage}
\caption{\label{fig:fig8} Generalised collisional-radiative recombination coefficients for ${\rm O}^{+4}+e \rightarrow {\rm
O}^{+3}$ (a) $ \scriptnew {ACD}(2\rms^2~^1S \rightarrow 2\rms^2 2\rmp~^2P)$  (b) $\scriptnew {ACD}(2\rms^2~^1S \rightarrow
2\rms 2\rmp^2~^4P)$ (c) $\scriptnew {ACD}(2\rms 2\rmp~^3P \rightarrow 2\rms 2\rmp^2~^4P)$}
\end{center}
\end{figure}
Figure \ref{fig:fig7} shows ${\scriptnew {PEC}}$ coefficients for the C II 858~\AA~spectrum line.  The coefficients depend on both
electron temperature and electron density in general.  A common practice in spectral analysis is to observe principally the strongest
resonance line of an ion.  Such emission is driven largely from the ground state (figure \ref{fig:fig7}a) and because of the large
A-value, the density sensitivity occurs at relatively high density.  Thus such resonance line emission at moderate to low densities
mostly reflects temperature and the distribution of the ionisation stages.  Comparison in near equilibrium ionisation balance plasmas
of line ratios from the same ion, by contrast, is mostly directed at electron density and relies on the presence of metastables and
spin changing collisional processes to confer the sensitivity.  As shown earlier, the balance of the dominant ground and metastable
populations is disturbed in dynamic plasmas and so density sensitivity may be modified by the dynamic state.  The distinction of the
metastable driven ${\scriptnew {PEC}}$ in $\scriptnew{GCR}$ modelling, as illustrated in figure \ref{fig:fig7}b, allows more complete
diagnostic study and the possibility of separation of the two effects.  In strongly recombining plasmas (most commonly photo-ionised
astrophysical plasmas) the direct contribution of recombination to the emission may dominate the excitation part.  The ${\scriptnew
{PEC}}^{(rec)}$, as illustrated in figure \ref{fig:fig7}c, is then required.  In the fusion context, multi-chordal spectral
observations are important for the study of impurity transport especially near sources.  Visible and quartz UV observations are
convenient and this places a requirement for ${\scriptnew {PEC}}$s from higher quantum shells.  It is this primarily which defines the
span of our {\it low-levels}  for population modelling.  Emission from higher $n$-shells is significantly affected by cascading from
yet higher levels.  The full machinery of projection as described in section \ref{sec:sec2.1} is necessary for our global ambition of
20\% precision for emissivity coefficients.                 

\begin{figure}[htp]
\begin{center}
   \begin{minipage}[t]{8.4cm}
   \raggedright
\ \psfig{file=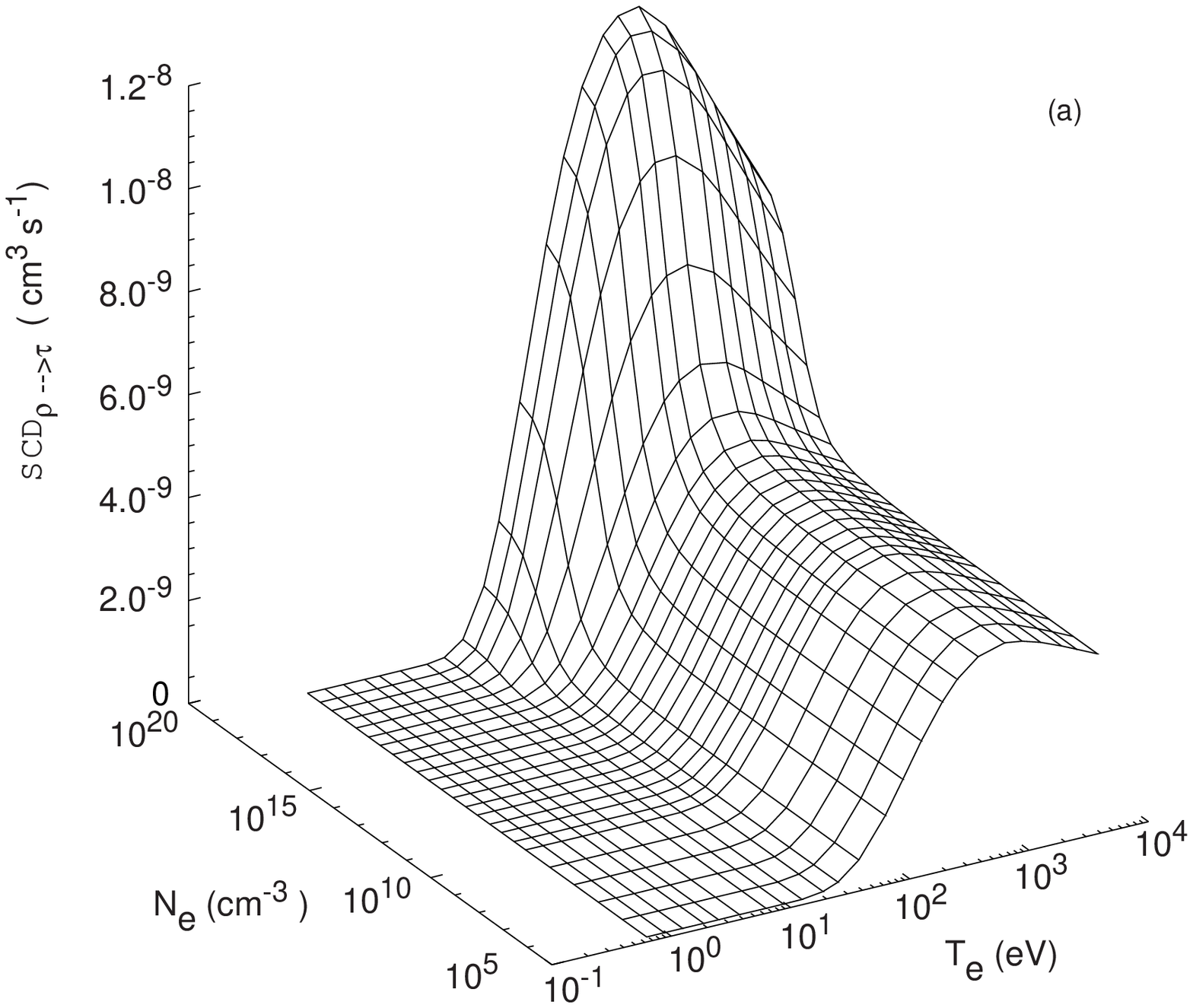,width=8.4cm}
\ \psfig{file=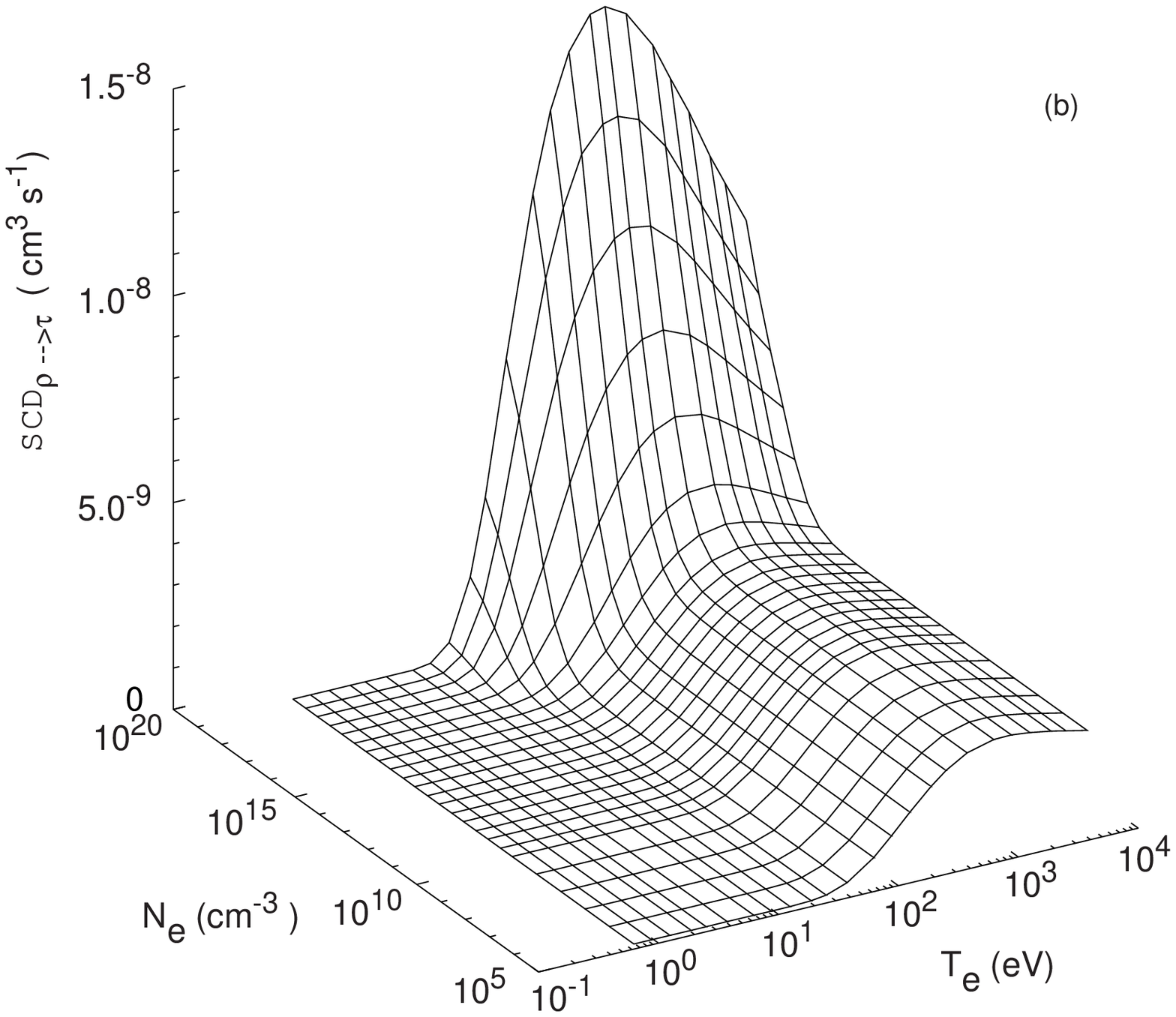,width=8.4cm}
   \end{minipage}
\caption{\label{fig:fig9} Generalised collisional-radiative ionisation coefficients for ${\rm O}^{+3}+e \rightarrow {\rm O}^{+4}
+e+e $ (a) $ \scriptnew {SCD}(2\rms^2 2\rmp~^2P \rightarrow 2\rms^2~^1S)$  (b) $\scriptnew {SCD}(2\rms^2 2\rmp~^2P \rightarrow
2\rms 2\rmp ~^3P)$ }
\end{center}
\end{figure}
Figure \ref{fig:fig8} illustrates the $\scriptnew{GCR}$ recombination coefficients.  At low electron density, radiative and dielectronic
recombination dominate.  For capture from metastables, alternate Auger branching can largely suppress the dielectronic part of
the surfaces. Thus figure \ref{fig:fig8}a shows the characteristic exponential rise at the temperature for excitation of the
main parent (core) transition of dielectronic recombination and then the subsequent fall-off (as $\sim T_e^{-3/2}$), in
contrast with \ref{fig:fig8}c.  At moderate densities, suppression of the high $n$-shell populations, principally populated by
dielectronic recombination, through re-ionisation occurs and the coefficient falls in the dielectronic recombination region.  At
very high density, three-body recombination becomes effective, preferentially beginning at the lower electron temperatures.  It
is evident that recombination is less effective from the metastable at relevant ionisation balance temperatures.  Models which
ignore the role of capture from the metastable parent (which may be the dominant population) can lead to substantial errors in
recombination coefficients, while simply excluding all capture from the metastable cannot deliver the precision sought for
current modelling.  For light elements in astrophysical plasmas, the zero-density coronal assumption for the recombination
coefficient is still frequently made.  This cannot be justified even at solar coronal densities.    

\begin{figure}[htp]
\begin{center}
   \begin{minipage}[t]{8.4cm}
   \raggedright
\ \psfig{file=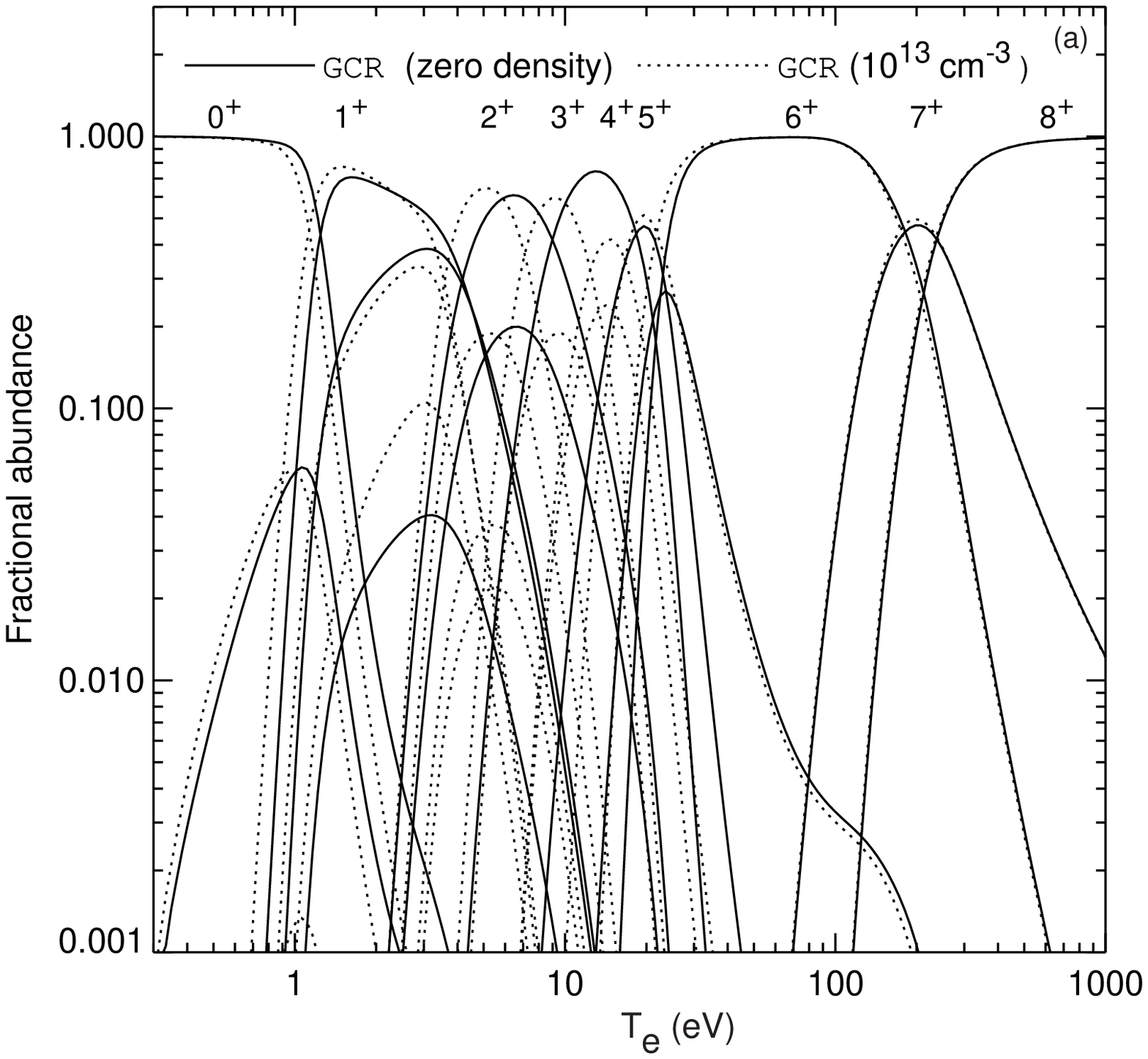,width=8.4cm}
\ \psfig{file=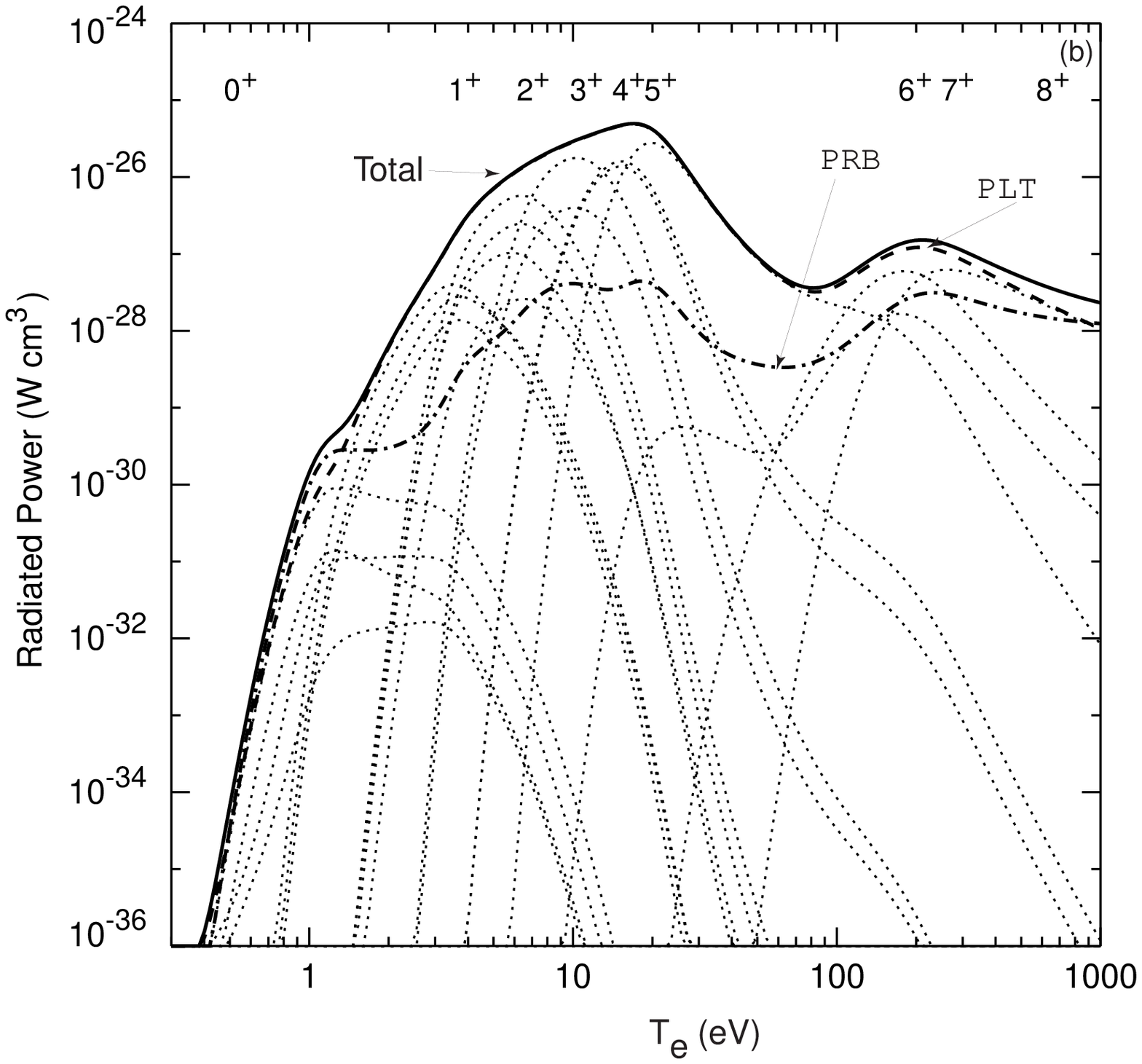,width=8.4cm}
\ \psfig{file=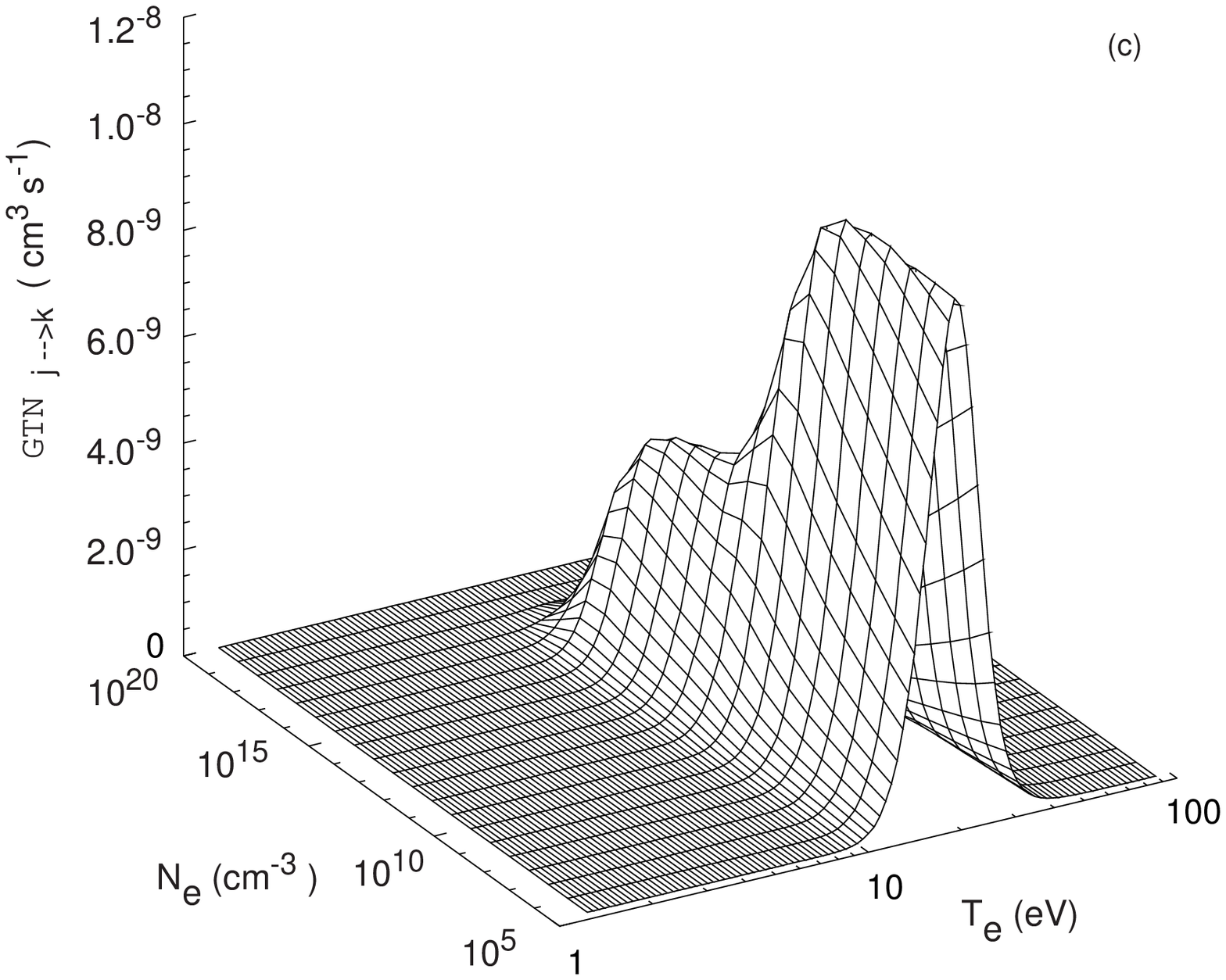,width=8.4cm}
   \end{minipage}
\caption{\label{fig:fig10} Equilibrium ionisation balance of oxygen (a) Equilibrium fractional abundances of metastables
from the generalised collisonal-radiative (b) The total radiated power function showing the components driven by the
different metastables (c) The $\scriptnew{GTN}$ function for the  O II 629.7 \AA~line showing the contributions parts
from the different metastables of the $z$-times and $(z+1)$-times ionised ions.}
\end{center}
\end{figure}
Figure \ref{fig:fig9} illustrates the $\scriptnew{GCR}$ ionisation coefficients. At low electron density, the coefficient is dominated
by direct ionisation, including excitation/auto-ionisation,  from the driving metastable.  Relatively high electron densities are required before
the stepwise contribution begins.  It is primarily the excitation to, and then further excitation and ionisation from, the first
excited levels which controls this.  The ground and metastable resolved coefficients both show the same broad behaviour as the usual
(stage to stage) collisonal-radiative coefficient, tending to a finite limit at very high density.  It is to be noted that the
$\scriptnew{XCD}$ coefficients are required to be able to construct a meaningful stage to stage collisional-radiative ionisation
coefficient from the generalised progenitors.  It remains the case that most plasma modelling (certainly in the fusion area) is not
adjusted to the use of the generalised coefficients as source terms.  Reconstruction of stage to stage source terms (at the price of a
reduction in modelling accuracy) from the generalised coefficients is still  is a requirement and is addressed more fully in section
\ref{sec:sec5}.        

The generalised coefficients may be used to establish the equilibrium ionisation balance for an element in which the
dominant ground and metastable populations are distinguished, that is the fractional abundances 
\begin{equation}
\label{eqn:eqn50}
\left ( \frac{N_{\sigma}^{[z]}}{N^{[tot]}} \right ) : \sigma=1,\cdots, M_z; z=0, \cdots , z_0 
\end{equation}
where $M_z$ is the number of metastables for ionisation stage $z$ and 
\begin{equation}
\label{eqn:eqn51}
N^{[tot]}=\sum_{z=0}^{z_0} N^{[z]}=\sum_{z=0}^{z_0}\sum_{\sigma=0}^{M_z} N_{\sigma}^{[z]}
\end{equation}
in equilibrium.  Writing $\underline{N}^{[z]}$ for the vector of populations $N_{\sigma}^{[z]}$, the equilibrium population fractions are obtained
from solution of the matrix equations.    
\begin{eqnarray}
\label{eqn:eqn52}
	\left[\begin{array}{llll}
		\underline{\scriptnew{C}}^{[0,0]} & N_e~\underline{\scriptnew{R}}^{[1 \rightarrow 0]} & 0 & . 	\\
		N_e\underline{\scriptnew{S}}^{[0 \rightarrow 1]} & \underline{\scriptnew{C}}^{[1,1]} &
		N_e~\underline{\scriptnew{R}}^{[2 \rightarrow 1]} & .	\\
		0 & N_e\underline{\scriptnew{S}}^{[1 \rightarrow 2]} & \underline{\scriptnew{C}}^{[2,2]} & .    \\
		0 & 0                                    & N_e\underline{\scriptnew{S}}^{[2 \rightarrow 3]} & . \\ 
		. & .  & . & . 
	      \end{array}
	 \right]
	\left[\begin{array}{l}
		\underline{N}^{[0]}  		\\
		\underline{N}^{[1]}   		\\
		\underline{N}^{[2]}   		\\
		\underline{N}^{[3]}   		\\
		.  
	       \end{array}
	 \right]_{equil}=0.
\end{eqnarray}
These in turn may be combined with the $\scriptnew{PLT}$ and  $\scriptnew{PRB}$ to obtain the equilibrium radiated power
loss function for the element as
\begin{eqnarray}
\label{eqn:eqn53}
P^{[tot]} & = & \sum_{z=0}^{z_0} P^{[z]}\left ( \frac{N^{[z]}}{N^{[tot]}} \right )_{equil} \nonumber \\
&  = & \sum_{z=0}^{z_0}\sum_{\sigma=0}^{M_z} ( \scriptnew{PLT}^{[z]}_{\sigma}+\scriptnew{PRB}^{[z]}_{\sigma})\left (
\frac{N^{[z]}_{\sigma}}{N^{[tot]}}\right )_{equil}.
\end{eqnarray}
The equilibrium fractional abundances and equilibrium radiated power function are illustrated in figure \ref{fig:fig10}. 
It is useful at this point to draw attention to emission functions which combine emission coefficients with equilibrium
fractional abundances.  They are commonly used in differential emission measure analysis of the solar atmosphere (Lanzafame
\etal, 2002) where
they are called $G(T_e)$ functions.  In solar astrophysics, it is assumed that the $G(T_e)$ are functions of the single
parameter $T_e$ (either from a zero-density coronal approximation or by specification at fixed density or pressure) and
usually the abundance of hydrogen relative to electrons $N_H/N_e$ is incorporated in the definition.  For finite density
plasmas, in the generalised collisonal-radiative picture, we define $\scriptnew{GTN}$ functions, parameterised by $T_e$
and $N_e$ as
\begin{eqnarray}
\label{eqn:eqn54}
\scriptnew{GTN}^{[z]}_{j \rightarrow k} &  = &  \sum_{\sigma}\scriptnew{PEC}^{[z](exc)}_{\sigma, j \rightarrow k}\left (
\frac{N_{\sigma}^{[z]}}{N^{[tot]}} \right )_{equil}
\nonumber \\
&& +\sum_{\nu'}\scriptnew{PEC}^{[z](rec)}_{\nu',j \rightarrow k}\left ( \frac{N_{\nu'}^{[z+1]}}{N^{[tot]}} \right )_{equil}. 
\end{eqnarray}
These are strongly peaked functions in $T_e$.  The precision of the present modelling and data, including the full
density dependence, is in principle sufficient to allow bivariate differential emission measure analysis (Judge \etal,
1997) although  this remains to be carried out. The general behaviour of a $\scriptnew{GTN}$ function is illustrated in
figure \ref{fig:fig10}c.       

\section{Computations and archiving derived data for applications}
\label{sec:sec5}
\begin{figure}[htp]
\begin{center}
   \begin{minipage}[t]{8.4cm}
   \raggedright
\ \psfig{file=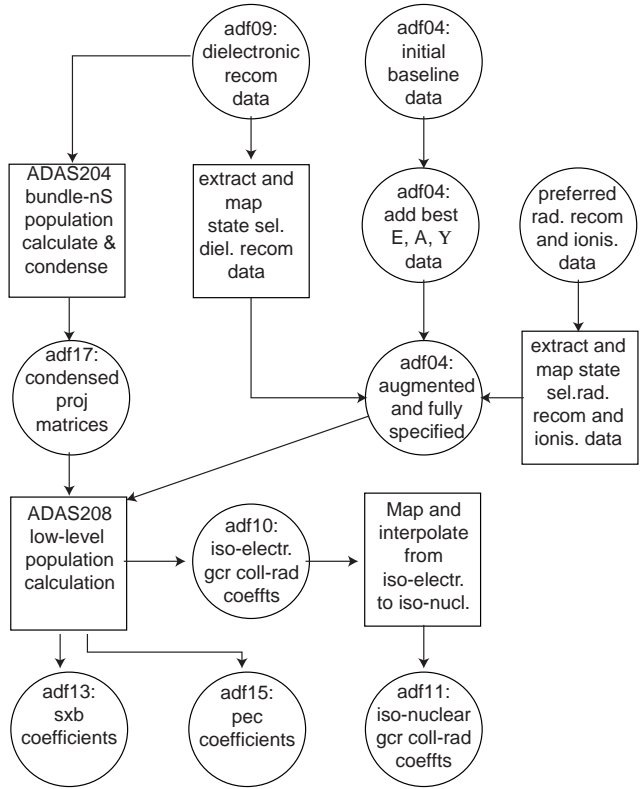,width=8.4cm}
   \end{minipage}
\caption{\label{fig:fig11} Schematic of computational steps in production of generalised collisonal-radiative data.  Data sets
are shown by circular outline and program elements by rectangles. The key members have been incorporated and assigned  names in
the ADAS system.}
\end{center}
\end{figure}
The organisation of the main calculations and data flow is shown in figure \ref{fig:fig11}.  The calculations executed for the
paper have been implemented in general purpose codes and attention has been given to the precise specification of all data sets
and the machinery for accessing and manipulating them.  This includes initial data, intermediate and driver data as well as the
final $\scriptnew{GCR}$ products and follows ADAS Project practice.  Population structure modelling requires an initial input dataset of
energy levels, transition probabilities and collisional rates of ADAS data format {\it adf04} which is complete for an
appropriate designated set of low levels.  For the light elements, best available data were assembled and verified as described
in section \ref{sec:sec3}.  These data are substitutes for more moderate quality, but complete, {\it baseline} data prepared
automatically (see {\it GCR - paper III}).  The {\it adf04} files are required for LS-terms.  In practice, we find it most
suitable to prepare the data for LSJ levels and then bundle back to terms.  For $\scriptnew{GCR}$ modelling, state selective
recombination and ionisation coefficients must be added to form the fully specified {\it adf04} file. The LS-coupled
dielectronic coefficients are mapped in from very large comprehensive tabulations of data format {\it adf09} prepared as part of
the {\it DR Project} (Badnell \etal,2003). The resulting {\it adf04} files are the ADAS preferred data sets and are available
for elements helium to neon. 

The two population codes, called ADAS204 and ADAS208, work together.  ADAS204 is the bundle-$nS$ model.  Data on the metastable parent structure, quantum
defects, auto-ionisation thresholds and autoionsation rates are required.  These data may be extracted from the
{\it adf04} files and it has been helpful to make the driver dataset preparation automatic.  High quality shell selective
dielectronic data are essential and this is part of the provision in the {\it adf09} files described in the previous paragraph. 
ADAS204 provides complete population solutions, but extracts from these solutions the condensed influence on the low n-shells,
as {\it projection matrices}, for connection with the calculations of ADAS208.  It is to be noted that the main on-going
development is refinement of atomic collision rates between the key low levels.  The projection matrices are not subject to
frequent change and so are suitable for long-term archiving ({\it adf17}).  ADAS208 is the low-level resolved population model
which delivers the final data for application.  It draws its key data from the fully configured {\it adf04} file and supplements
these with projection data.  The evaluation of the population structure takes place at an extended set of $z$-scaled electron
temperatures and densities (see section \ref{sec:sec2}) and this means that the resulting $\scriptnew{GCR}$ coefficient data are suited
to interpolation along iso-electronic sequences.  Thus the initial tabulation of $\scriptnew{GCR}$  coefficients is in iso-electronic
datasets.  It is convenient to implement the gathering and mapping from iso-electronic to iso-nuclear in a separate step,
which also supports the merging back to the unresolved stage-to-stage picture if required.  The production of $\scriptnew{PEC}$
and  $\scriptnew{SXB}$ coefficients is directly to iso-nuclear oriented collections.  The number of $\scriptnew{PEC}$s from the
population calculations can, in principle, be very large.  We restrict these by a threshold magnitude and to
particular important spectral regions.  It is straightforward to rerun ADAS208 to generate $\scriptnew{PEC}$ or
$\scriptnew{SXB}$ coefficients alone in spectral intervals of one's choice.  The separation of the ADAS204 and ADAS208 tasks and
the ease of modifying data within an {\it adf04} file means that `what-if' studies on the sensitivity of  the derived data to
fundamental data uncertainly can readily carried out.  Special ADAS codes enable detailed study of cumulative error and
dominating sources of uncertainty such that error surfaces, for example for a $\scriptnew{PEC}$ as a function of electron
temperature and density, may be generated.  Such error (uncertainty) analysis for theoretical derived coefficients and its
utilisation in the confrontation with diagnostic experiments is the subject of a separate work (see O'Mullane \etal, 2005).                        

\section{Conclusions}
\label{section:conclusions}

The requirements for precise modelling of spectral emission and the relating of ionisation stages in thermal plasmas
have been considered.  Collisional-radiative methodologies have been developed and extended to enable the full role of
metastables to be realised, so that this generalised ($\scriptnew{GCR}$) picture applies to most dynamically evolving
plasmas occurring in magnetic confinement fusion and astrophysics.

The procedures are valid up to high densities.  The studies presented in the paper explore the density effects in detail
within the $\scriptnew{GCR}$ picture and show that the density dependencies of excited ion populations and of effective
rate coefficients cannot be ignored.

Specific results are presented for light elements up to neon and the computations are carried out in an atomic basis of
terms (LS-coupled).  Such modelling will remain sufficient up to about the element argon, beyond which a level basis
(intermediate coupling) becomes necessary.  Heavier elements will be examined in further papers of this series.  

Considerable attention has been given to the generation and assembly of high quality fundamental data in support of the 
$\scriptnew{GCR}$ modelling.  Also datasets of fundamental and derived data have been specified precisely and codes have
been organised following the principles of the ADAS Project.  The product of the study is the preferred ADAS data for the
light element ions at this time.                
  
\section{Acknowledgements}

The Atomic Data and Analysis Structure, ADAS, was originally developed at JET Joint Undertaking.  
   
%\begin{thebibliography}{}
\section*{References}
\begin{harvard}

   \item [] Bates D R, Kingston A E and McWhirter R W P 1962 {\it Proc. Roy. Soc. A} {\bf 267} 297.
   \item [] Badnell N R 1986 {\it J. Phys. B} {\bf 19} 3827.
   \item [] Badnell N R  and Pindzola M S 1989 {\it Phys. Rev. A} {\bf 39} 1685.
   \item [] Badnell N R 1997 {\it J. Phys. B} {\bf 30} 1.
   \item [] Badnell N R, O'Mullane M G, Summers H P, Altun Z, Bautista M A, Colgan J, Gorczyca T W,
            Mitnik D M, Pindzola M S and Zatsarinny O 2003 {\it Astron. \& Astrophys.} {\bf 406} 1151.
   \item [] Ballance C P, Badnell N R, Griffin D C, Loch S D and Mitnik D 2003 {\it J. Phys. B} {\bf 36} 235.
   \item [] Bartschat K and Bray I 1996 {\it J. Phys. B} {\bf 29} L577.
   \item [] Bray I and Stelbovics A T 1993 {\it Phys. Rev. Lett.} {\bf 70} 746.
   \item [] Bryans P, Torney M, Paton I D, O'Mullane M G, Summers H P, Whiteford A D, Bingham R and
            Kellett B J 2005 {\it Plasma Physics \& Control. Fusion} - submitted.      
   \item [] Burgess A and Summers H P 1969 {\it Astrophys. J.} {\bf 157} 1007.
   \item [] Burgess A and Summers H P 1976 {\it Mon. Mot. R. Astr. Soc.} {\bf 174} 345.
   \item [] Burgess A and Summers H P 1987 {\it Mon. Mot. R. Astr. Soc.} {\bf 226} 257.
   \item [] Burgess A and Tully J A  1992 {\it Astron. \& Astrophys.} {\bf 254} 436.
   \item [] Colgan J, Loch S D, Pindzola M S, Ballance C P and Griffin D C 2003 {\it Phys. Rev. A} {\bf 68}:032712.
   \item [] Cowan R D 1981 {\it The Theory of Atomic Structure and Spectra}
                   (University of California Press: Berkeley)
   \item [] Froese Fischer C, Godefroid M and Olsen J 1997 {\it J. Phy. B} {\bf 30} 1163.
   \item [] Froese Fischer C, Gaigalas G and Godefroid M 1997 {\it J. Phys. B} {\bf 30} 3333.
   \item [] Fleming J, Vaeck N, Hibbert A, Bell K L and Godefroid M R 1996a {\it Phys. Scr.} {\bf 53} 446.
   \item [] Fleming J, Bell K L, Hibbert A, Vaeck N and Godefroid M R 1996b {\it Mon. Not. R. Astr. Soc.} {\bf 279} 1289.
   \item [] Fritsche S and Grant I P 1994 {\it Phys. Scr.} {\bf 50} 473.
   \item [] J\"{o}nsson P, Froese Fischer C and Tr\"{a}bert E 1998 {\it J. Phys. B} {\bf 31} 3497.
   \item [] Judge P G, Hubeny V and Brown J C 1997 {\it Astrophys. J.} {\bf 475} 275.
   \item [] Kelly R L 1987 {\it J Phys Chem Ref Data} {\bf 16} Suppl 1. 
   \item [] Lanzafame A C, Brooks D H, Lang J, Summers H P, Thomas R J and Thompson A M 2002 {\it Astron. \& Astrophys.} {\bf  384}
	    242.
   \item [] Loch S D, Colgan J, Pindzola M S, Westermann M, Scheuermann F, Aichele K, Hathiramani D and
            Salzborn E 2003 {\it Phys. Rev. A} {\bf 67}:042714 
   \item [] Loch S D, Witthoeft M, Pindzola M S, Bray I, Fursa D V, Fogle M, Schuch R and Glans P
            2005 {\it Phys. Rev. A} {\bf 71}:012716 
   \item [] McWhirter R W P and Summers H P 1984 {\it Applied Atomic Collision Physics Vol 2: Fusion} chap. 3 (Ed.
            Harrison, M.E.)(Acad. Press, New York). 
   \item [] NIST Standard Reference Database 61, Database for Atomic Spectroscopy version 1.
   \item [] Nussbaumer H O and Storey P J 1979 {\it Astron. \& Astrophys.} {\bf 74} 244.
   \item [] O'Mullane M G, Summers H P, Whiteford A D 2005 {\it Plasma Physics \& Control. Fusion} - to be submitted       
   \item [] Opacity Project Team 1995 {\it Opacity Project} Vol 1, Institute of Physics Publishing, Bristol, UK.
   \item [] Pindzola M S and Badnell N R 1990 {\it Phys. Rev. A} {\bf 42} 6526.
   \item [] Pindzola M, Griffin D C and Bottcher C 1986 {\it Atomic Processes in Electron-Ion and Ion
			      Collisions}, NATO Advanced Study Institute, series B: vol. 145 (ed. Brouillard, F.) ( Plenum, New
			      York).
   \item [] Pindzola M and Robicheaux F 1996 {\it Phys. Rev. A} {\bf 54} 142.
   \item [] Ramsbottom C A, Berrington K A and Bell K L 1995 {\it At. Data and Nucl. Data Tables} {\bf 61} 105.
   \item [] Sampson D H, Goett S J and Clark R E M 1984 {\it At. Data and Nucl. Data Tables} {\bf 30} 125.
   \item [] Sampson D H and Zhang H L 1988 {\it Phys. Rev. A} {\bf 37} 3765.
   \item [] Summers H P 1974 {\it Mon. Mot. R. Astr. Soc.} {\bf 169} 663.
   \item [] Summers H P 1977 {\it Mon. Mot. R. Astr. Soc.} {\bf 178} 101.
   \item [] Summers H P and Hooper M H 1983 {\it Plasma Physics} {\bf 25} 1311. 
   \item [] Summers H P 1993 {\it JET Joint Undertaking Report} JET-IR(93)07.
   \item [] Summers H P 2004 {\it The ADAS User Manual, version 2.6} http://adas.phys.strath.ac.uk.
   \item [] Tully J A, Seaton M J and Berrington K A 1991 {\it J. Phys. B} {\bf 23} 381.
   \item [] Wiese W L, Fuhr J R and Deters T M 1996 {\it J. Phys. Chem. Ref. Data} {\bf 7}. 
   \item [] Whiteford A D, Badnell N R, Barnsley R, Coffey I H,  O'Mullane M G, Summers H P and Zastrow K-D 2005 {\it X-ray
            Diagnostics of Astrophysical Plasmas} AIP Conf. Proc. (Ed. Smith, R.) - in press.
    
\end{harvard}
%\end{thebibliography}

\end{document}